\documentclass[manuscript]{aastex}
\pdfoutput=1

\newcommand{\ie}{{\it i.e.}}    
\newcommand{\etal}{et al.}      
\newcommand{\inv}{^{-1}}
\newcommand{\kms}{\mathrm{km~s}\inv}
\newcommand{\ergsex}{\mathrm{~erg~s}\inv\mathrm{~cm}^{-2}}
\newcommand{\HST}{\emph{HST}}

\newcommand{\Nii}{[\ion{N}{2}]}
\newcommand{\Sii}{[\ion{S}{2}]}
\newcommand{\Oii}{[\ion{O}{2}]}
\newcommand{\Oiii}{[\ion{O}{3}]}
\newcommand{\Hii}{\ion{H}{2}}
\newcommand{\Neiii}{[\ion{Ne}{3}]}
\newcommand{\Halpha}{H$\alpha$}
\newcommand{\Hbeta}{H$\beta$}

\newcommand{\Hdelta}{H$\delta$}
\newcommand{\Caii}{\ion{Ca}{2}}
\newcommand{\Mgii}{\ion{Mg}{2}}

\shorttitle{TKRS2}
\shortauthors{Wirth et al.}

\begin{document}

\title{The Team Keck Redshift Survey 2: MOSFIRE Spectroscopy of the
  GOODS-North Field}

\author{Gregory D. Wirth\altaffilmark{2,3}, 
  Jonathan R. Trump\altaffilmark{4,5,6}, 
  Guillermo Barro\altaffilmark{4},
  Yicheng Guo\altaffilmark{4},
  David C. Koo\altaffilmark{4},
  Fengshan Liu\altaffilmark{4},
  Marc Kassis\altaffilmark{2}, 
  Jim Lyke\altaffilmark{2}, 
  Luca Rizzi\altaffilmark{2}, 
  Randy Campbell\altaffilmark{2}, 
  Robert W. Goodrich\altaffilmark{2,7}, 
  \and
  S. M. Faber\altaffilmark{4}}

\altaffiltext{1}{ Based in part on data obtained at the W.~M. Keck
  Observatory, which operates as a scientific partnership among the
  California Institute of Technology, the University of California and
  the National Aeronautics and Space Administration. The generous
  financial support of the W.~M. Keck Foundation made the Observatory
  possible.}
\altaffiltext{2}{W. M. Keck Observatory, 65-1120
  Mamalahoa Hwy, Kamuela, HI 96743}  
\altaffiltext{3}{Current address: NEON, Inc., 1685 38th St., Suite 100, Boulder, CO, 80301; gregory.wirth@gmail.com}
\altaffiltext{4}{University of
  California Observatories, Department of Astronomy and Astrophysics,
  University of California, Santa Cruz, CA 95064}
\altaffiltext{5}{Department of Astronomy and Astrophysics, 525 Davey
  Lab, The Pennsylvania State University, University Park, PA 16802}
\altaffiltext{6}{Hubble Fellow}
\altaffiltext{7}{Current address: GMTO Corp., 6628 N. Muscatel Ave., San Gabriel, CA, 91775-1534}

\begin{abstract}

  We present the Team Keck Redshift Survey 2 (TKRS2), a near-infrared
  spectral observing program targeting selected galaxies within the
  CANDELS subsection of the GOODS-North Field.  The TKRS2 program
  exploits the unique capabilities of MOSFIRE, an infrared
  multi-object spectrometer which entered service on the Keck~I
  telescope in 2012 and contributes substantially to the study of
  galaxy spectral features at redshifts inaccessible to optical
  spectrographs.  The TKRS2 project targets 97 galaxies drawn from
  samples that include $z\approx2$ emission-line galaxies with
  features observable in the $JHK$ bands as well as lower-redshift
  targets with features in the $Y$ band.  We present a detailed
  measurement of MOSFIRE's sensitivity as a function of wavelength,
  including the effects of telluric features across the $YJHK$
  filters.  The largest utility of our survey is in providing
  rest-frame-optical emission lines for $z>1$ galaxies, and we
  demonstrate that the ratios of strong, optical emission lines of $z
  \approx 2$ galaxies suggest the presence of either higher N/O
  abundances than are found in $z\approx0$ galaxies or low-metallicity
  gas ionized by an active galactic nucleus.  We have released all
  TKRS2 data products into the public domain to allow researchers
  access to representative raw and reduced MOSFIRE spectra.
  
\end{abstract}


\section{Introduction}\label{sxn:intro}

The peak of the cosmic star-formation rate (SFR) occurs at $z \approx
2$ \citep[e.g.,][]{mad14}, but the star-formation processes occurring
in galaxies observed at that epoch differ significantly from those
occurring in galaxies seen today.  In contrast to galaxies such as the
Milky Way which form stars continuously in giant molecular clouds,
$z>1$ galaxies may be dominated by discrete and recurrent starbursts
\citep[e.g.,][]{pap05}.  The dynamics of most $z>1$ galaxies are also
frequently disordered \citep{kas12}, exhibiting clumps of
high-pressure gas or young stars \citep[e.g.,][]{cow95,
  elm04,guo12,guo15}, unlike the smoother, rotation-dominated disks of
massive, star-forming galaxies observed at the present epoch.
Reconciling the disparate properties of present-day galaxies with
those observed at $z>1$ remains an active and crucial research topic
for tracking galaxy evolution.

Rest-frame optical spectroscopy offers a powerful means to probe the
physical conditions of galaxies by potentially revealing the
evolutionary pathways at play for $z \approx 2$ galaxies.  In addition
to measuring redshifts needed for deriving basic galaxy properties
such as light and stellar mass, rest-frame-optical spectra provide
emission lines that probe the SFR \citep[e.g.,][]{ken98}, gas-phase
metallicity \citep[e.g.,][]{kew01}, and nuclear activity of galaxies
\citep[e.g.,][]{bpt81, vo87, kew06}.  At $z>1$, rest-frame-optical
features are redshifted into the near infrared, a wavelength regime in
which observations were once limited to single-slit spectrographs; thus,
completing near-IR redshift surveys has historically been
prohibitively time-intensive.  The advent of multi-object, near-IR
spectrographs on 8--10~m-class telescopes has recently enabled several
new surveys of rest-frame-optical emission lines in $z>1$ galaxies
\citep[e.g.,][]{yos10, 3dhst, kas13, tru13, stei14, mosdef, wis15}.

In this work, we present the Team Keck Redshift Survey 2 (TKRS2), a
spectroscopic survey of 97 distant galaxies exploiting the unique
capabilities of the MOSFIRE spectrometer on the Keck~I telescope.
Section \ref{sxn:mosfire} describes the key characteristics enabling
MOSFIRE to complete infrared spectroscopic surveys of the type
previously limited to optical spectrometers.  In \S\ref{sxn:selection}
we describe the mask design and target selection, and \S\ref{sxn:spec}
details the spectroscopic observations, data reduction, and method for
determining redshifts.  The redshift catalog appears in
\S\ref{sxn:catalog}.  We present some basic analyses of the
spectroscopic data in \S\ref{sxn:analysis}, including a
characterization of the MOSFIRE sensitivity across the $YJHK$ filters,
comparison to previous redshifts, and a discussion of the galaxies'
rest-frame optical emission-line ratios.

\section{The MOSFIRE Spectrometer}\label{sxn:mosfire}

Our survey relies on the unique capabilities of the W. M. Keck
Observatory's newest instrument, the Multi-Object Spectrometer For
Infra-Red Exploration \citep[MOSFIRE,][]{mcl12}.  MOSFIRE combines
flexible multi-slit spectroscopy capability with high throughput,
making it the ideal near-infrared instrument for studying the faintest
and most distant galaxies.  MOSFIRE features a $2048\times2048$ pixel
HAWAII-2RG HgCdTe detector array from Teledyne Imaging Sensors that
couples high quantum efficiency with low noise and low dark current.
The operating range of 0.97--2.41~$\micron$ covers the $YJHK$ infrared
passbands, with wavelength coverage of 0.97--1.12~\micron\ in $Y$,
1.15--1.35~\micron\ in $J$, 1.47--1.80~\micron\ in $H$, and
1.95--2.40~\micron\ in $K$.  Observers can acquire spectra in any one
of these passbands by setting the diffraction grating and
order-sorting filters appropriately, although simultaneous
observations in multiple passbands \citep[cf. VLT's
XSHOOTER;][]{ver11} are not possible.  The resolving power for the
default slit width of $0\farcs7$ is $R$ = 3,380 in $Y$, 3,310 in $J$,
3,660 in $H$, and 3,620 in $K$, corresponding to FWHM spectral
resolutions of 3.1~\AA\ in $Y$, 3.7~\AA\ in $J$, 4.4~\AA\ in $H$, and
6.0~\AA\ in $K$.

The feature which most distinguishes MOSFIRE from other slit
spectrographs --- optical as well as infrared --- is the configurable
slit mechanism residing within its cryogenic dewar.  Previous IR
multi-slit spectrometers have relied on custom-milled slitmasks that
required thermal cycling of the dewar for installation and removal.
In contrast, the MOSFIRE Configurable Slit Unit (CSU) is a
fully-robotic mechanism which is remotely controlled and allows the
instrument to remain at cryogenic operating temperature throughout
each observing run, yielding observations of greater stability.  The
CSU mechanism puts 92 pairs of matching bars in the re-imaged telescope
focal plane, forming up to 46 separate slits of length $7\farcs0$ that
can be positioned independently within the $6\farcm1\times6\farcm1$
MOSFIRE field of view, or can be combined to form longer slits.  The
MOSFIRE mask design software includes a target prioritization scheme,
and the sky density of our faint ($H\lesssim24$) galaxy targets
normally resulted in 25--30 targets being assigned to slits on each
mask.  The width of each slit can be precisely and independently
adjusted based on the needs of the observing program; however, we
employed a consistent slit width of $0\farcs7$ in the present survey
to balance the competing desires for improved resolution (dictating
narrower slits) and throughput (requiring wider slits).  MOSFIRE
observers can easily adjust slit widths during the course of a night
to account for variations in seeing conditions, although a slit width
of $0\farcs7$ is generally a good match to the customary excellent
($\sim0\farcs5$) Maunakea seeing in the near-IR.

The MOSFIRE CSU also enables a novel mask alignment technique.  Many
telescopes --- Keck~I and II included --- cannot be positioned to
better than the standard slit width based on telescope encoders or
guider-based alignment techniques, nor can most instruments be set to
the precisely correct rotator angle based on positional feedback
alone.  The common strategy for aligning a multi-slit layout with the
corresponding (faint) sky targets is to include two or more
``alignment boxes'' on a slitmask; each box is placed at the expected
location of a star with accurately known astrometric position
\citep[e.g.,][]{kassis12}.  When observers select alignment stars to
be much brighter than the faint science targets, short images acquired
with the mask in place show the locations of the stars relative to the
edges of each alignment box, thereby indicating the rotation and
decentering of the mask relative to the alignment stars (and, thus, to
the science targets).  Ideally, small adjustments of the telescope
pointing and instrument rotator angle by appropriate amounts lead to
optimal alignment within a couple of iterations.  Dedicating space on
the mask for alignment boxes does, however, sacrifice area that could
be devoted to science slits.

In contrast to this customary method, the procedure of aligning
MOSFIRE's slits with the corresponding sky targets exploits the CSU's
ability to reposition the slits in real time.  When setting up on a
given field, the observer commands the CSU to recast several
(generally 3--4) of the bar pairs to form ``alignment boxes'' that are
actually just extra-wide slits (normally $4\arcsec\times7\arcsec$).
After obtaining alignment between these alignment boxes and their
corresponding stars, the observer commands the bar pairs to return to
their science configuration to form slits that will collect spectra of
science targets.  As a result, the full complement of MOSFIRE slits
can be devoted to science observations.  Given the benefits that the
MOSFIRE CSU offers over traditional slitmask spectroscopy, we
anticipate that future optical and IR spectrometers will exploit this
technology.

\section{Target Selection \& Mask Design}
\label{sxn:selection}

\begin{figure}
  \epsscale{.90}
  \plotone{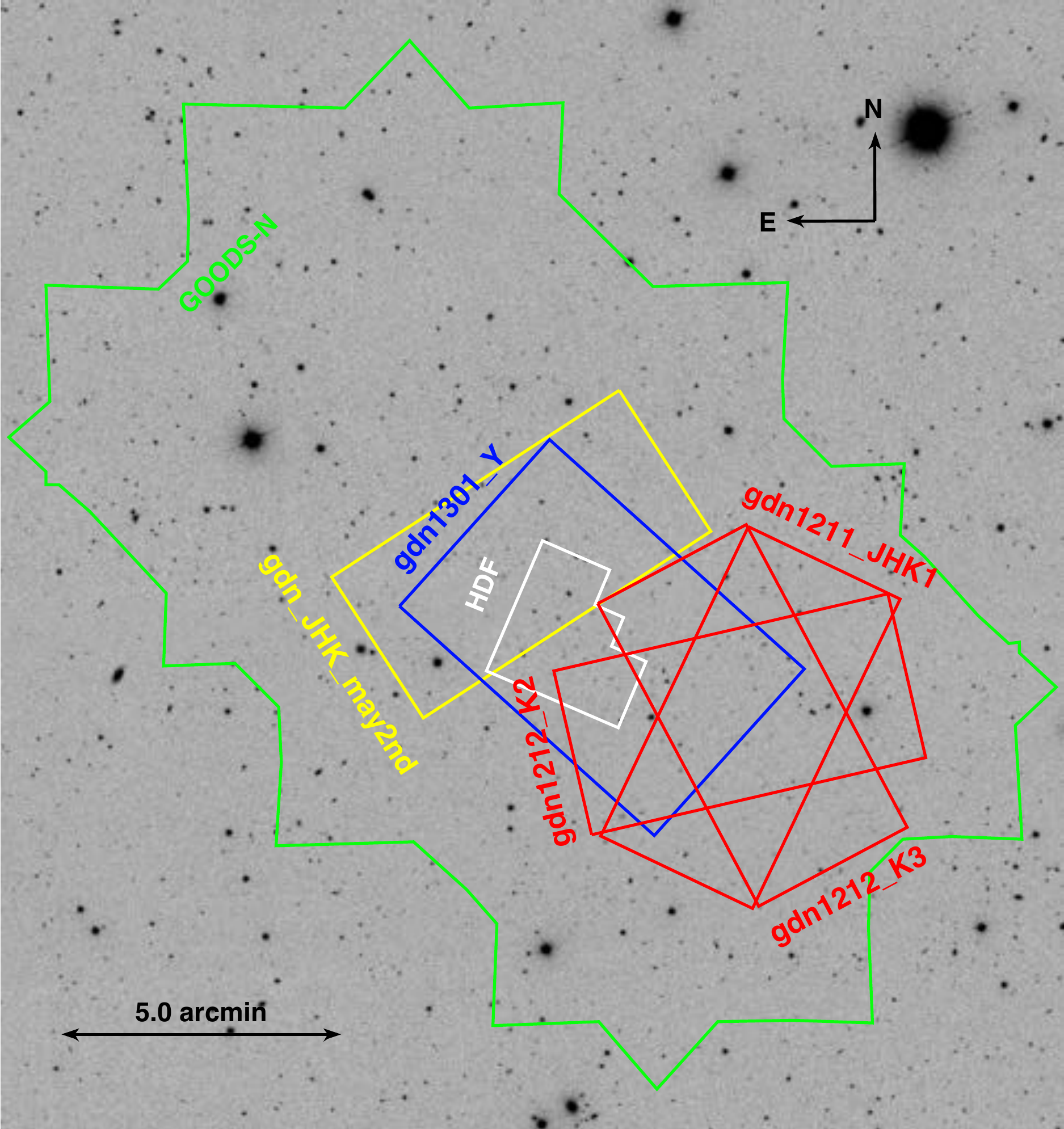}
  \caption{
    \label{fig:fields}
    TKRS2 survey fields shown against an image of the GOODS-North
    field obtained from the Sloan Digital Sky Survey
    \cite[SDSS,][]{sdss}.  The small white polygon indicates the
    approximate boundary of the HDF-North imaging survey region, and
    the green polygon denotes the GOODS-North ACS imaging survey
    region.  The red, yellow, and blue rectangles indicate the fields
    of the TKRS2 survey.  The image is oriented with north up and east
    to the left, with angular scale as indicated by the bar at lower
    left.  }
\end{figure}

We designed the TKRS2 project to test the capabilities of MOSFIRE on
diverse categories of extragalactic sources.  Our survey targets the
south-central region of the GOODS-North survey field \citep[see
  Fig.~\ref{fig:fields};][]{gia04}, an area in which the CANDELS program
\citep{candels} has compiled a superb set of complementary data from
the \emph{Hubble Space Telescope (HST)} and other observatories
operating in regimes from radio to X ray.  We obtained
reset-frame-optical spectra for a sample of galaxies in the redshift
range $0.5<z<2.5$, gathering observations in all four of MOSFIRE's
$YJHK$ filters with varying exposure times.  Nearly all (90 of 97) of
the sources are galaxies without previous high-resolution near-IR
spectroscopy.  We describe each target category below.

\begin{figure}
 \epsscale{.80}
 \plotone{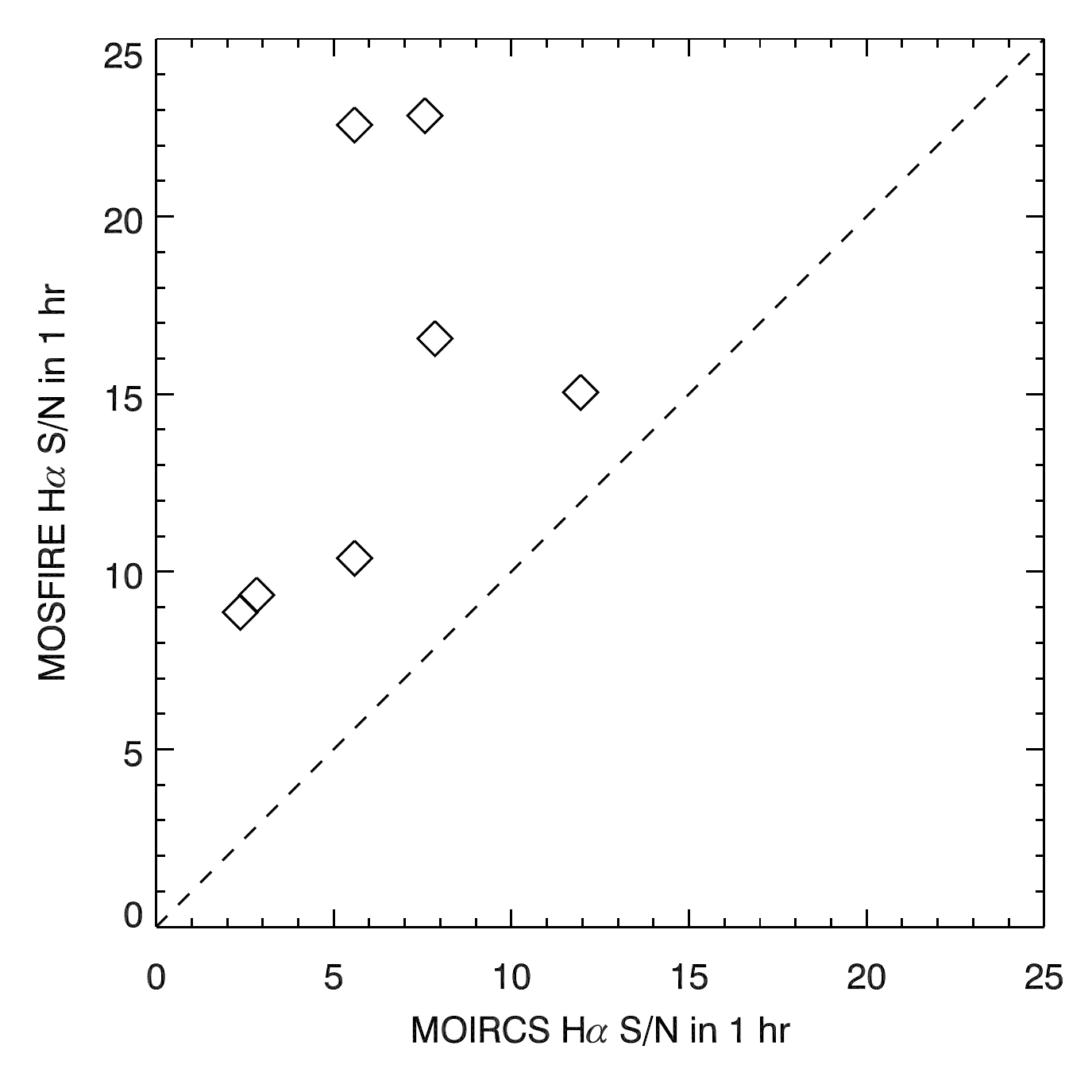}
  \caption{
    \label{fig:moircscompare}
    A comparison of the emission line S/N achieved by MOSFIRE and
    MOIRCS in 1~h.  The MOIRCS \Halpha\ line fluxes and errors are
    taken from Table 3 of \citet{yos10}.  MOSFIRE achieves $\sim$2--3
    times higher S/N as MOIRCS in the same exposure time, with
    some scatter due to presence of of telluric features in
      some spectra.}
\end{figure}

\begin{itemize}
  
\item \textit{emline}: 83 galaxies expected to show emission lines
  (\Oii\ $\lambda\lambda 3726,3729$~\AA, \Hbeta\ $\lambda4861$~\AA,
  \Oiii\ $\lambda\lambda4959,5007$~\AA, \Halpha\ $\lambda6563$~\AA,
  \Nii\ $\lambda\lambda6548,6583$~\AA, or
  \Sii\ $\lambda\lambda6717,6731$~\AA) at the appropriate redshifts to
  be observable in the $Y$, $J$, $H$, or $K$ bands. Two thirds of
  these sources have prior spectroscopic redshifts from optical
  surveys \citep{tkrs,red06,bar08,fer09,coo11} or estimated redshifts
  derived from low-resolution, near-IR \HST/WFC3 G141 grism
  observations (B.~J. Weiner, P.I.).  Observing these targets with
  MOSFIRE adds new, rest-frame-optical emission lines useful for
  characterizing the physical gas conditions of these galaxies.  The
  remainder have photometric redshifts based on the CANDELS
  multi-wavelength catalog (Barro \etal, in preparation); we included
  such objects on the masks at substantially lower priority.  We also
  selected extended and clumpy galaxies to test the ability of the
  routinely excellent Maunakea seeing coupled with with MOSFIRE's high
  spectral resolution to generate resolved kinematic measurements.
  
\item \textit{moircs}: seven galaxies, each of which had previous
  Subaru+MOIRCS $H$ or $K$ observations \citep{yos10}.   We included
  these targets primarily to test the performance and efficiency of
  MOSFIRE compared to MOIRCS in $H$ and $K$ (see
  Fig.~\ref{fig:moircscompare}), but we also observed in other
  passbands to access additional emission lines. The ``moircs''
  galaxies are similar to the bright galaxies in the
  ``emline'' category.
  
\item \textit{quiescent}: seven galaxies having spectral energy
  distributions (SEDs) indicative of quiescent stellar populations
  (\ie, with low specific SFR) and intended as targets for $Y$-band
  spectroscopy.  We selected these galaxies based on their 
  SFR and $U-V$ vs.\ $V-J$ rest-frame colors ($UVJ$
  criterion; e.g., \citealt{wil09}).  We also required the quiescent
  galaxy targets to have photometric redshifts of $0.9 < z < 1.3$ in
  order to test the capability of MOSFIRE for measuring absorption
  lines (e.g., \Caii~H$+$K $\lambda$4000~\AA, \Hdelta, \Hbeta\ and
  \Mgii~B~$\lambda$5178~\AA) in the $Y$ band.  In general, the 1~h
  depth of the observations resulted in only tentative absorption line
  detections, requiring deeper observations to confirm.
  Interestingly, the spectra of 3 quiescent galaxies include weak
  emission lines, consistent with (in each case) a weak MIPS 24~$\mu$m
  flux, a weak X-ray detection, and a nearby star-forming neighbor.

\end{itemize}

We designed spectroscopic masks for observations in the $Y$, $J$, $H$,
and $K$ MOSFIRE observing bands.  On the $Y$ mask, we assigned highest
priority to the low-density ``quiescent'' galaxies.
We then filled the remaining slits on the mask with ``emline''
galaxies having spectroscopic redshifts (from TKRS and \citet{bar08}) in
the range $0.5<z<0.7$, such that \Halpha, \Nii, and/or \Sii\ are
visible in the $Y$ band.  The $Y$ mask included 32 targets, a high
target density given the maximum of 46 slits available on MOSFIRE.

We designed two masks to observe in each of $J$, $H$, and $K$.  We
designed an additional two masks for $K$ observations only, with
different position angles (by 40--$80\degr$) such that at least one
mask would place a slit within $40\degr$ of a galaxy's kinematic axis.
Kinematic analysis of these multi-orient slit observations will be
presented in future work (Simons \etal, in preparation).  In selecting
specific targets, we gave highest priority to the ``moircs'' galaxies,
and assigned ``emline'' sources in the redshift range $2.1<z<2.6$ to
the remaining slits.  At these redshifts, \Oii\ lines lie in $J$,
\Hbeta\ and \Oiii\ in $H$, and \Halpha, \Nii, and \Sii\ in $K$.
About two thirds of the targeted emission-line galaxies have redshifts
derived from low-resolution \HST/WFC3 G141 grism observations (B.~J. Weiner,
P.I.).  We selected these sources over those having only photometric
redshifts, and gave slightly higher priority to galaxies with
visibly extended and clumpy morphologies \citep[based on imaging from
  the CANDELS project;][]{candels, candels2} in order to test the
capabilities of MOSFIRE for studying internal galaxy kinematics.

The next sections discuss the observations and data reduction of
the survey. Overall, the survey achieved high redshift
completeness ($>80\%$) for emission-line galaxies (``emline'' and
``moircs'' categories) but had marginal success on quiescent targets.

\section{Spectroscopy}\label{sxn:spec}

\subsection{Observations}

Former WMKO Director Taft Armandroff generously contributed several
nights of his Director's discretionary observing time allocation
toward this campaign.  Its principal aims were to complement
our previous optical survey of the GOODS-North field \citep{tkrs},
demonstrate the capabilities of the new MOSFIRE instrument on the
Keck~I telescope, and establish a public resource by releasing all
data products from the survey to the astronomical community.  We
employed MOSFIRE to acquire spectra in the GOODS-North field over a
series of partial nights spanning the period from November 2012 to May
2013.  Table~\ref{tab:obs} summarizes the observations.

To facilitate accurate subtraction of sky and instrumental background
emission, spectroscopic observing sequences consisted of a series of
dithered exposures alternating between two positions symmetrically
offset from the initial pointing.  We employed individual exposure
times totaling 180~s on-sky integration time in the $Y$ and $K$ bands;
given the relatively greater temporal instability of telluric emission
in the $J$ and $H$ bands, we reduced integration times at these
wavelengths to 120~s.  To confirm that the slits remained well aligned
with the faint science targets throughout exposure sequences generally
lasting an hour or more, mask designs generally included one slit
intended to acquire the spectrum of a star which was significantly
brighter than the target galaxies.  The relatively bright (high S/N)
stellar spectrum has multiple applications: monitoring the
transparency of the sky; indicating the accuracy of the frame-to-frame
telescope dithers along the slit; tracking any pointing drifts
during the exposures along the slit direction; recording the variable
atmospheric absorptions after normalizing the continuum to a best-fit
stellar spectral type; and tracking the pointing offsets perpendicular
to the slit by determining the offsets between the telluric absorption
wavelengths to that of the telluric emission lines that fill the
slit. The data reduction adopted for this work as described in
\S\ref{sxn:dataReduction} only exploits the information on the dithers
along the slit. More advanced reductions \citep[see, e.g.,][]{mosdef}
are possible but beyond the scope of this release.

In keeping with common MOSFIRE practices, we acquired a standard
series of dome flatfield exposures in each passband in order to
characterize the instrumental response.  In the $K$ band, we also
acquired dome exposures with no dome-flat illumination, thus isolating
the thermal component of the dome emission.  Given the prevalence of
strong telluric emission in the $Y$, $J$, and $H$ bands which serve to
calibrate the wavelength scale, we acquired no arc lamp exposures in
these passbands.  We also obtained arc lamp exposures for wavelength
determination in the $K$ band due to the relatively weaker telluric
emission features at longer wavelengths.

\subsection{Data Reduction}
\label{sxn:dataReduction}

We processed all images using the \textsc{MosfireDRP} data reduction
pipeline\footnote{\url{http://www2.keck.hawaii.edu/inst/mosfire/drp.html}}
written by the MOSFIRE instrument development team and generously
shared with the observing community.  The pipeline is specifically
designed to accommodate dithered spectra.  The reduction procedure
virtually eliminates contamination from telluric emission lines by
independently combining observations from each of the two pointings,
scaling the combined images by the exposure time, shifting the images
by the dither amount, and computing the difference.  By default, the
pipeline derives the wavelength solution for the two-dimensional (2-D)
spectra using only the telluric emission lines.  We also experimented
with using arc lamp lines to derive wavelength solutions in the $K$
band, which has comparatively weaker telluric emission features than
the shorter-wavelength bands.  Using the arc lamp data for calibration
resulted in identical wavelength solutions, and so we retained the
standard wavelength calibration using telluric lines in our final
reduced spectra.

The end result of the \textsc{MosfireDRP} pipeline is a
sky-subtracted, wavelength-calibrated, rectified 2-D spectrum for each
slit.  To create one-dimensional (1-D) spectra, we used custom software
to fit a Gaussian function to the wavelength-collapsed image profile,
constraining the peak to lie within $\pm5$ pixels of the expected
object position.  We extracted the 1-D spectra from a
inverse-variance-weighted co-add within a boxcar window centered on the
best-fit peak and with a width $1.5\times$ the Gaussian
full-width-half-maximum (FWHM).  Within a single MOSFIRE filter, the 2-D
spectral trace is tilted by $\lesssim 1$ pixel, and so the flat trace
used by our extraction method results in reasonable 1-D spectra.

\subsection{Redshift Determination}
\label{sxn:determination}

To estimate the redshifts of the targets, at least two team members
independently used the \textsc{Specpro} software package
\citep{specpro} to inspect each of the MOSFIRE spectra.  This software
allows the user to fit various template spectra to the observed
spectra interactively.  We made use of the prior redshift information
for each target from optical spectroscopy
\citep{tkrs,red06,bar08,fer09,coo11} and IR grism spectroscopy
(B.~J. Weiner, P.I.), along with the estimated ``photo-z'' redshifts derived
from multiband photometry (Barro \etal, in preparation).

Reviewers recorded both the derived MOSFIRE redshift, $z_M$, and a
redshift quality parameter, $Q_M$, which denotes the reviewer's
confidence in the redshift estimate.  Values of $Q_M$ are as follows:

\begin{itemize}
  \item 0: no reported redshift due to lack of identifiable features;
  \item 1: speculative redshift based on a single line which is faint
    and/or blended with a sky feature;
  \item 2: ambiguous redshift based on a single line that does not
    match the photometric redshift;
  \item 2.5: ambiguous redshift based on a single line that matches
    (within $\Delta z/z<0.15$) the prior photometric redshift;
  \item 3: secure ($P>95\%$) redshift, typically including one strong
    emission feature and one or more additional weak features; and,
  \item 4: highly secure ($P>99\%$) redshift, generally exhibiting
    multiple strong emission features.
\end{itemize}

After reviewing all targets, we collated the results, re-inspected
each target for which the derived redshift or quality code differed,
and reached consensus on a final redshift and quality code.
Table~\ref{tab:zdist} indicates the number of galaxies receiving each
classification.

\section{Redshift Catalog}
\label{sxn:catalog}

We present the results of our survey in Table~\ref{tab:zcat} and on
the website\footnote{\url{http://arcoiris.ucsc.edu/TKRS2/}} devoted to the
survey. In addition to the target identification and position, the
table lists the class to which the target belongs and the apparent AB
magnitude of the target in \HST\ imaging.  Observational data from the
TKRS2 survey includes the list of MOSFIRE passbands in which we
observed the target, the total MOSFIRE exposure time devoted to the
target, and which spectral features we identified in the MOSFIRE
spectra.  Finally, several sources of redshift information appear in
the table, including the presumed redshift of the target from previous
spectroscopic surveys (when available) and the source of that prior
redshift, the estimated redshift of the source derived via multiband
photometry from the CANDELS survey, the redshift derived from the
present survey (accounting for prior information), and the redshift
quality code, $Q_M$, as described in \S\ref{sxn:determination}.

\begin{figure}
  \epsscale{1.0}
  \plotone{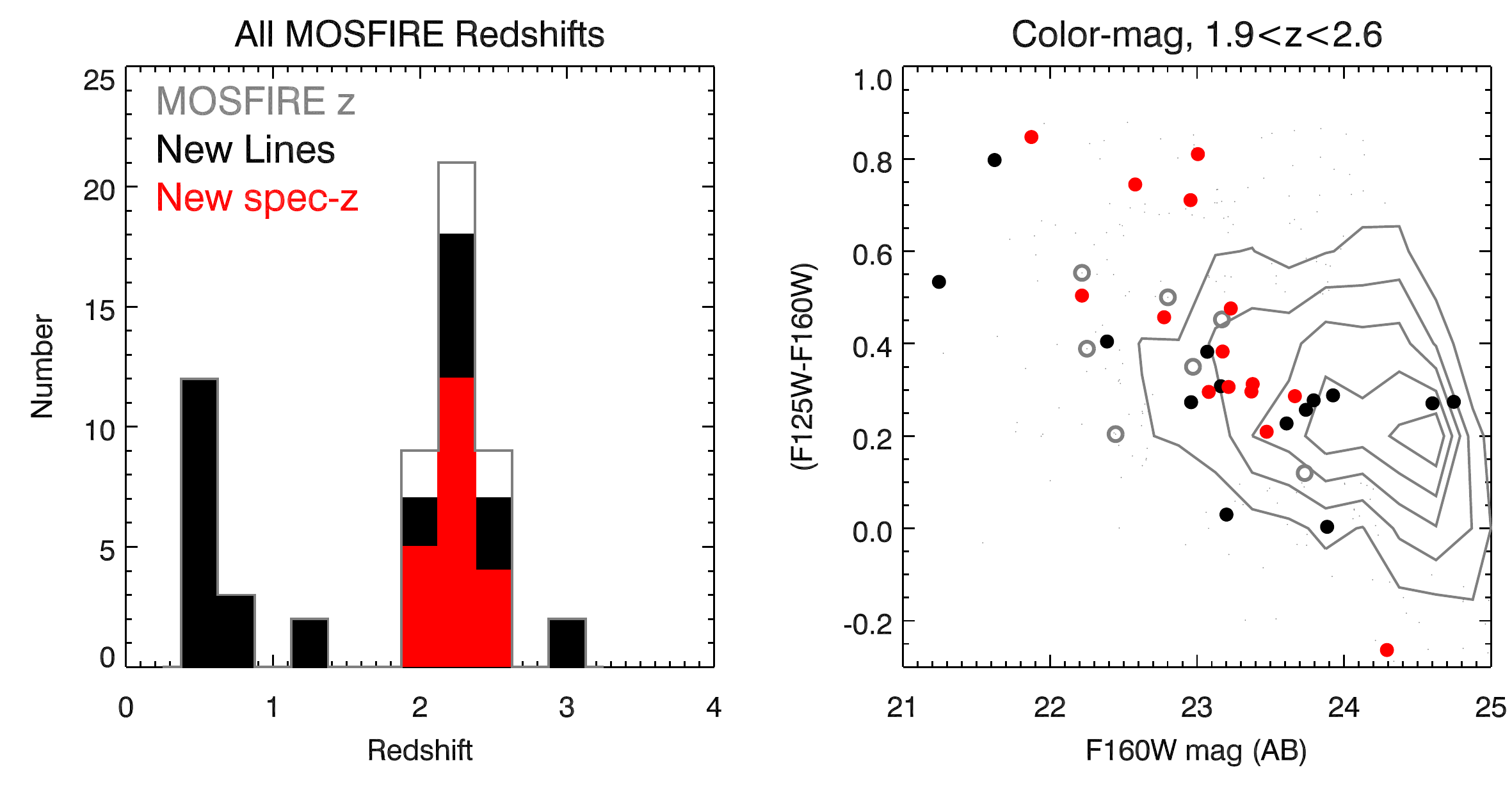}
  \caption{
    \label{fig:zhist}
    At left, a histogram showing the distribution of spectroscopic
    redshifts from our program.  The open histogram shows the full set
    of high-confidence ($Q_M \ge 3$) MOSFIRE redshifts.  The filled
    red histogram shows the set of targets without previous redshifts
    from optical spectroscopy, and the black filled histogram shows
    galaxies with new rest-frame optical emission-line measurements.
    At right, the $J-H$ vs.\ $H$ color-magnitude diagram for galaxies
    with MOSFIRE redshifts of $1.9<z<2.6$ (circles) and the larger
    galaxy population with photometric redshifts in the same range
    (gray contours and points).  Gray circles indicate galaxies with
    previous MOIRCS redshifts, filled red circles denote targets with
    new spectroscopic redshifts, and filled black circles designate
    galaxies with new rest-frame optical lines.  Our sources are
    more likely to have $H<24$ than the larger population, but
    otherwise span the same range of $J-H$ colors (roughly
    corresponding to rest-frame $U-B$ in this redshift range).}
\end{figure}

The distribution of high-quality ($Q_M \ge 3$) redshifts from our
survey is shown in Fig.~\ref{fig:zhist} (left panel).  For nearly
all of the galaxies with high-quality redshifts, our program provides
the first spectra with rest-frame optical emission lines.  The only
sources with previous near-IR spectroscopy (of moderate resolution)
are the seven ``moircs'' targets, which we included for comparison
of Keck+MOSFIRE with Subaru+MOIRCS; see \S\ref{sxn:compare}.  The
right panel of Fig.~\ref{fig:zhist} presents a color-magnitude diagram
comparing our $1.9<z<2.6$ galaxies to the larger population of
galaxies with photometric redshifts in the same range.  Galaxies with
high-quality ($Q_M \ge 3$) MOSFIRE redshifts tend to have $H<24$, but
are otherwise representative of the color distribution for the larger
population.

\begin{figure}
  \epsscale{1.0}
  \plotone{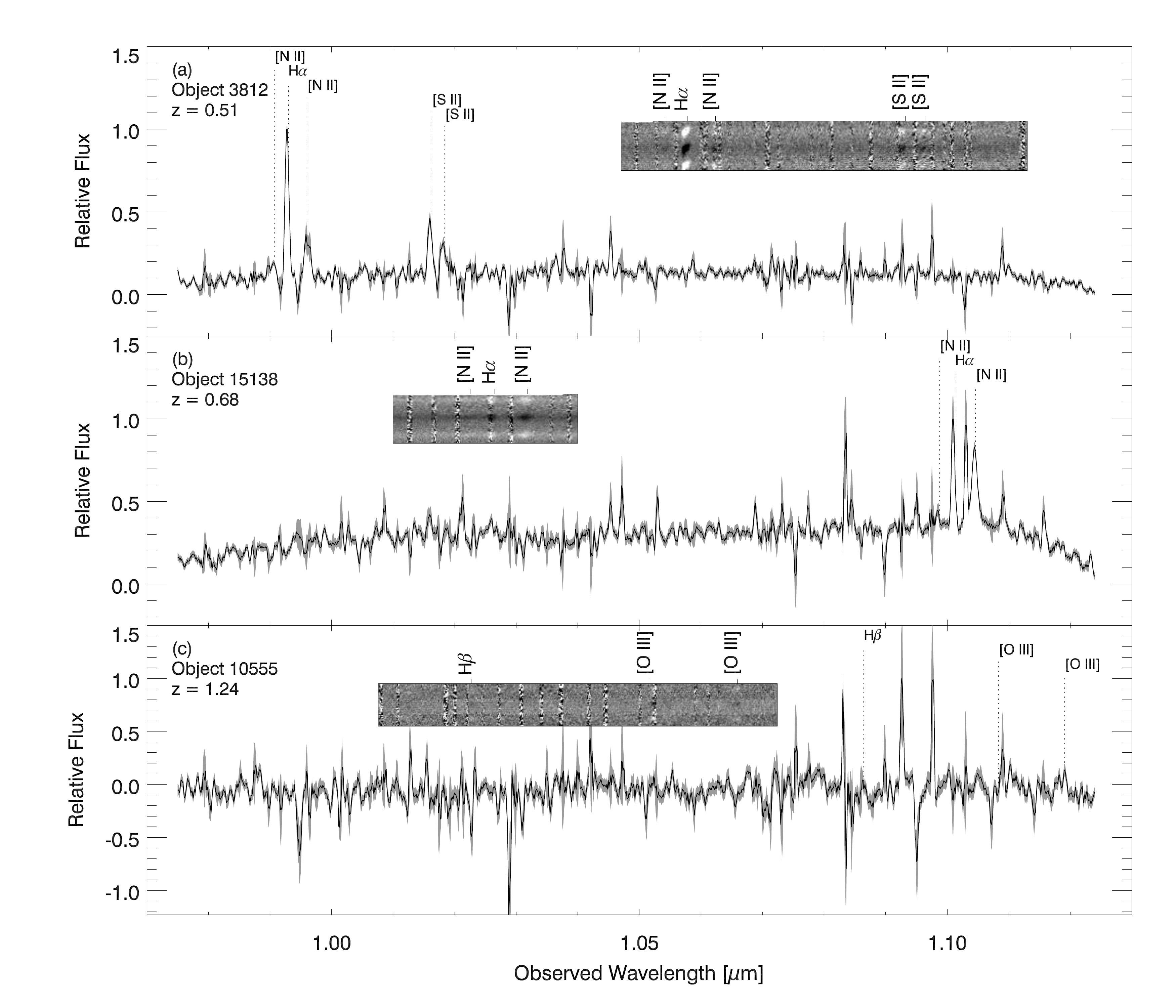}
  \caption{
    \label{fig:galleryY}
    Representative MOSFIRE $Y$-band (near-IR) spectra.  Shaded region
    indicates $\pm1\sigma$ uncertainty in the spectrum; labels
    indicate expected locations of notable emission-line features.  We
    applied five-pixel boxcar smoothing to 1-D spectra.  Insets show a
    portion of the corresponding 2-D spectrum to illustrate key
    emission features, with expected locations of features marked.
    (a)~Object 3812 is a typical $z\approx0.5$ galaxy showing \Halpha,
    \Nii, and \Sii\ emission, all apparent in the 2-D inset.
    (b)~Object 15138, an X-ray source, lies at somewhat higher
    redshift; inset shows \Halpha\ and \Nii.  (c)~Object 10555 at
    $z=1.24$ has faintly visible \Oiii\ $\lambda$5007~\AA\ and
    \Hbeta\ features, apparent in the inset; a terrestrial emission
    line obscures the \Oiii\ $\lambda$4959~\AA\ feature.  Confirming
    optical spectra provide confidence in the redshift despite the
    relatively low S/N.}
\end{figure}

\begin{figure}
  \epsscale{1.0}
   \plotone{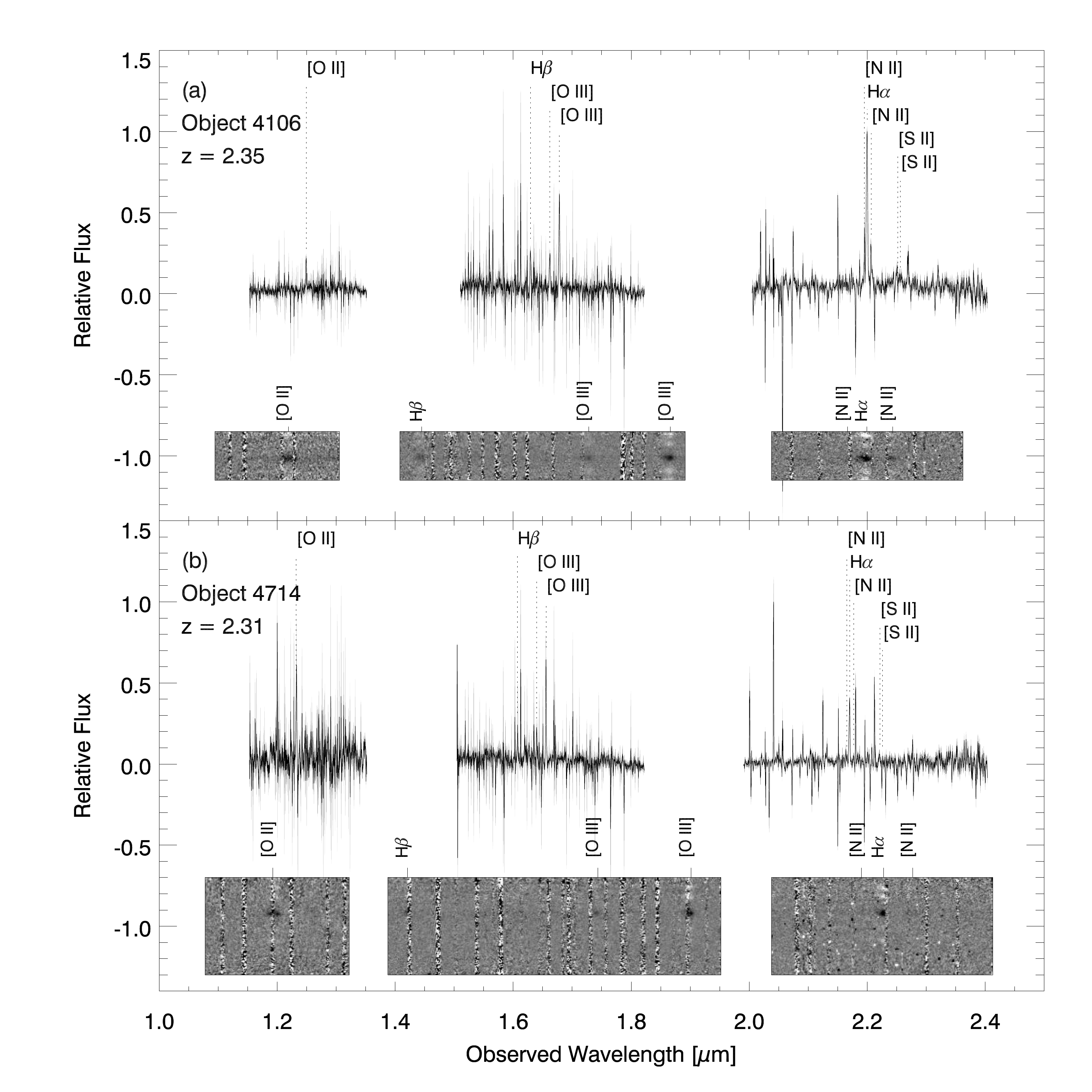}
  \caption{
    \label{fig:galleryJHK}
    Spectra representative of observations acquired in MOSFIRE's $JHK$
    passbands.  As in Fig.~\ref{fig:galleryY}, we applied five-pixel
    boxcar smoothing to supress random noise in the 1-D spectra.
    (a)~The spectrum of object 4106 (an X-ray source) shows
    \Oii\ visible in the MOSFIRE $J$ band, \Hbeta\ and \Oiii\ in $H$,
    as well as \Halpha, \Nii, and \Sii\ emission in $K$.  (b)~Object
    4714 is similar but with relatively fainter features.}
\end{figure}

\begin{figure}
  \epsscale{1.0}
  \plotone{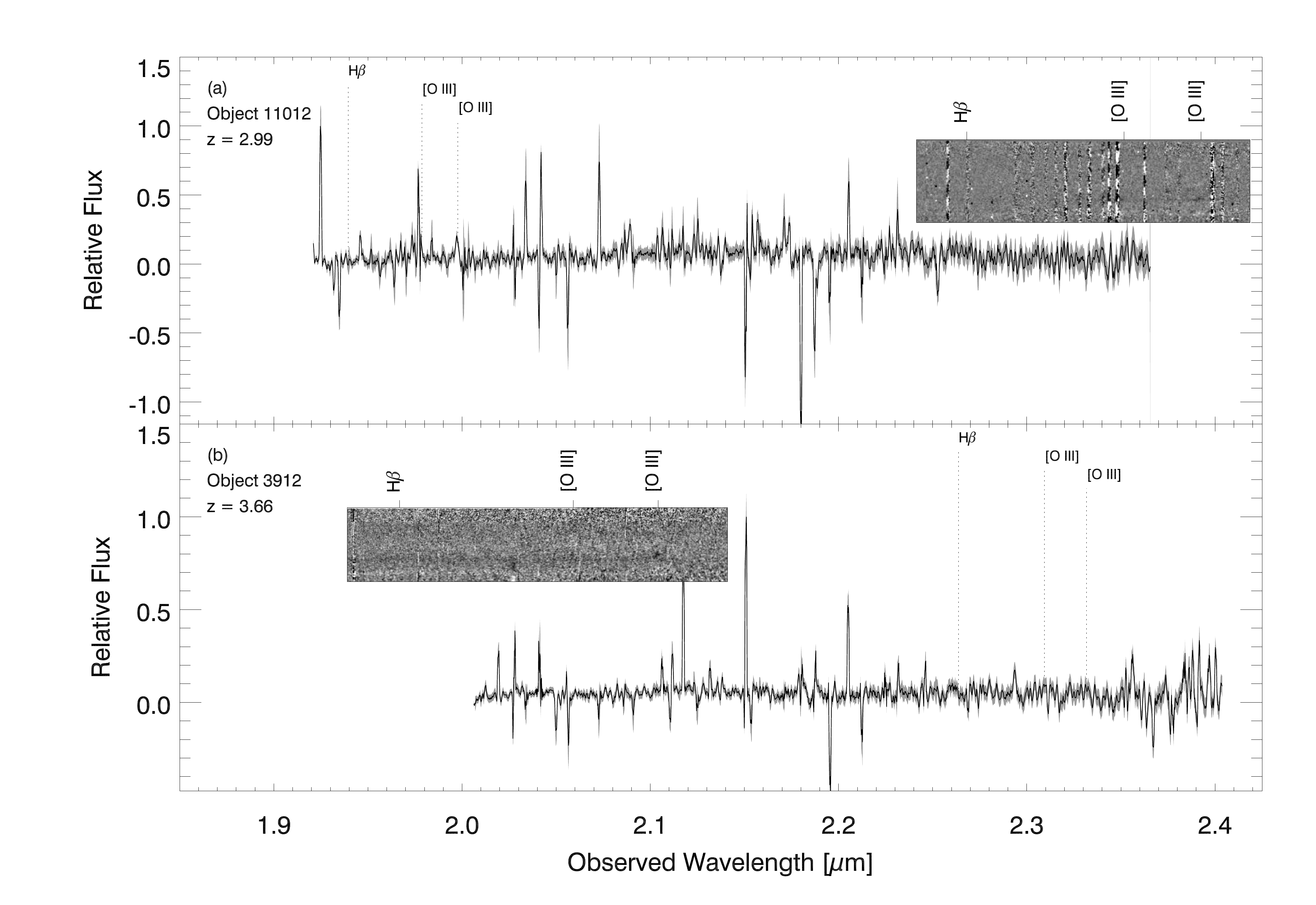}
  \caption{
    \label{fig:galleryK}
    Spectra representative of observations of high-redshift targets
    observed in MOSFIRE's $K$ passband.  As in
    Fig.~\ref{fig:galleryY}, we applied five-pixel boxcar smoothing to
    supress random noise in the 1-D spectra.  (a)~Object 11012 is a
    redshift $z=2.99$ X-ray source, though it has only weak emission
    lines.  (b)~The $z=3.66$ redshift of object 3912 is confirmed on
    the basis of the \Oiii\ $\lambda$5007~\AA\ emission line faintly
    visible in the 2-D inset.}
\end{figure}

\begin{figure}
 \epsscale{0.8}
 \plotone{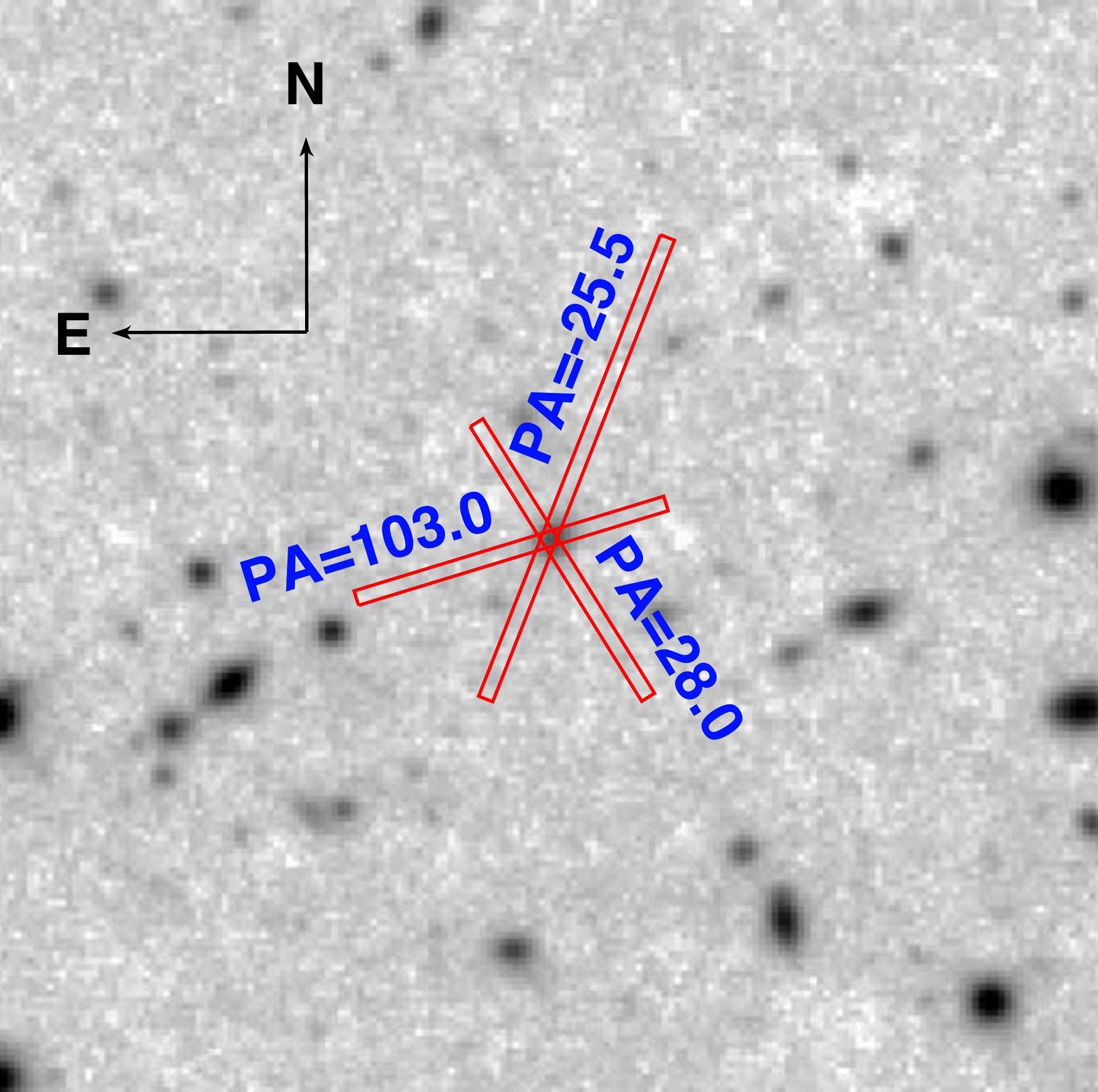}
  \caption{
    \label{fig:multiPA}
    Illustration of our survey strategy to observe selected targets at
    multiple position angles.  The figure shows a ground-based
    $K$-band image of object 4925, an emission-line galaxy at
    $z=2.245$.  Superimposed on this image are the projected sky
    locations of the MOSFIRE slit at three position angles,
    corresponding to three of our mask designs. }
\end{figure}

\begin{figure}
 \epsscale{0.75}
 \plotone{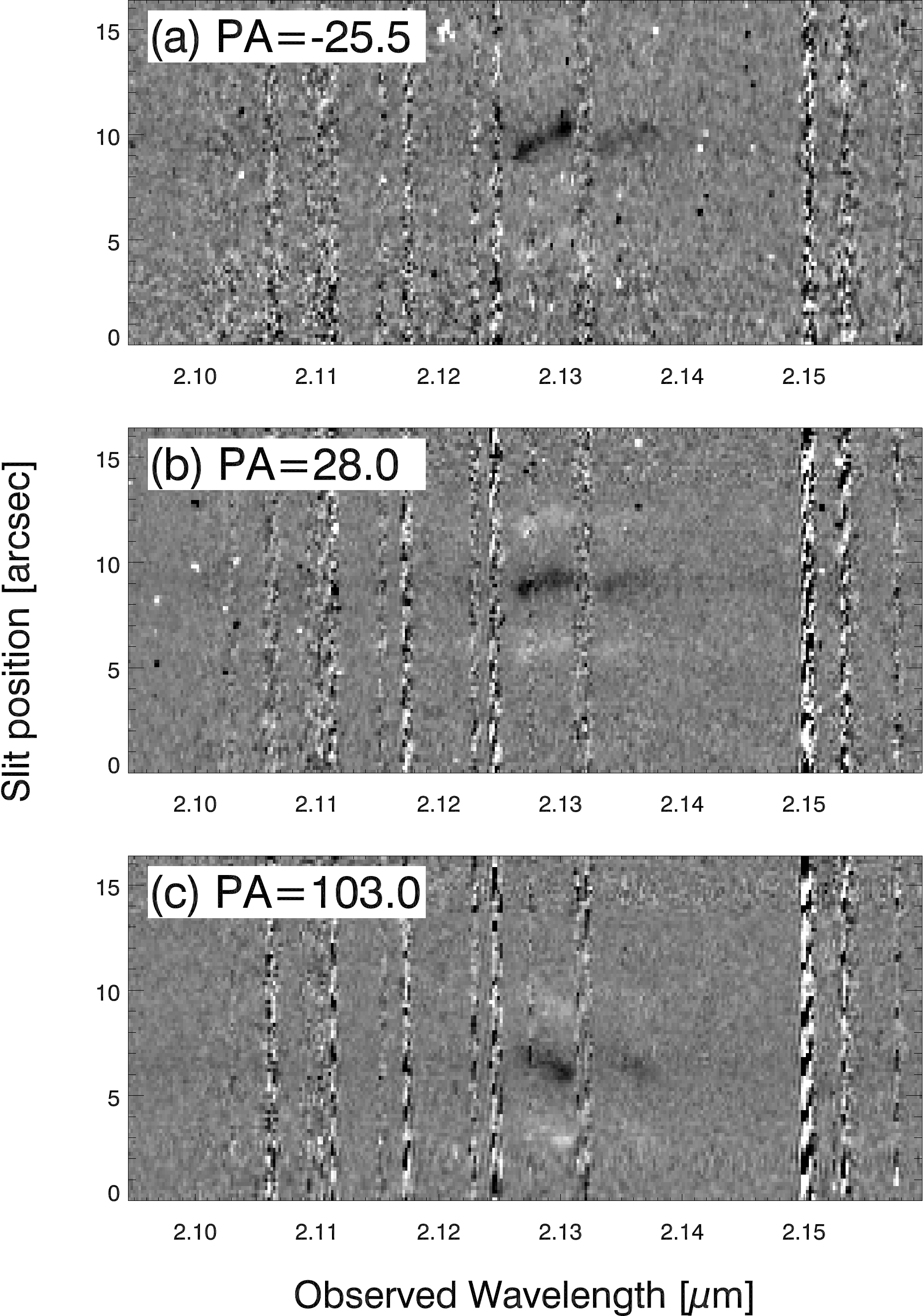}
  \caption{
    \label{fig:multiPA2}
    $K$-band MOSFIRE spectra we obtained at each of the three slit
    position angles indicated in Fig.~\ref{fig:multiPA}.  Visible in
    the spectra are the emission lines
    \Halpha\ $\lambda$6563~\AA\ (redshifted to $2.130~\micron$) and
    \Nii\ $\lambda$6583~\AA\ (observed at $2.136~\micron$).  Note that
    the velocity profile of the lines changes with position angle,
    allowing detailed investigation of the rotational properties of
    the target.  }
\end{figure}

Representative spectra from the survey appear in
Fig.~\ref{fig:galleryY}, depicting the $Y$ band,
Fig.~\ref{fig:galleryJHK}, displaying $JHK$, and
Fig.~\ref{fig:galleryK}, presenting $K$-only spectra.  Additionally,
Figs.~\ref{fig:multiPA} and \ref{fig:multiPA2} illustrate the benefit
of observing the same target at multiple position angles in order to
gain information about the rotational properties of distant galaxies.

\section{Analysis}\label{sxn:analysis}

\subsection{MOSFIRE Sensitivity}

We characterize the sensitivity of MOSFIRE as the flux limit to detect
an emission line at the 3$\sigma$ level in 1~h of on-target
integration.  The flux limit is not constant with wavelength, but
instead depends strongly on the presence of telluric emission
features.  For each filter, we empirically measure the pixel-by-pixel
noise as the normalized median absolute deviation (NMAD) from the set
of all continuum-subtracted spectra taken in each filter.  The noise
spectrum is converted from detector units to flux density using the
detector response function (available on the Keck MOSFIRE throughput
webpage\footnote{\url{http://www2.keck.hawaii.edu/inst/mosfire/throughput.html}})
and a conversion between $e^{-}~\mathrm{s}\inv$ and flux density
measured from twelve $z \approx 1.5$ galaxies with the same emission
lines measured by both MOSFIRE and the \HST/WFC3 G141 grism.  By
flux-calibrating with the WFC3 slitless grism we implicitly include a
slit loss correction (for galaxies of similar angular size as the
twelve $z \approx 1.5$ galaxies).  The slit loss correction is
typically a factor of 1.5--1.7 \citep{mosdef}.  We then convolve the
noise with a Gaussian, assuming a redshifted \Halpha\ emission line
with rest-frame width $\sigma=85~\kms$ ($\mathrm{FWHM}=200~\kms$).
This process results in the noise of the fiducial Gaussian emission
line centered at each pixel.

\begin{figure}   
 \epsscale{1.0}
 \plotone{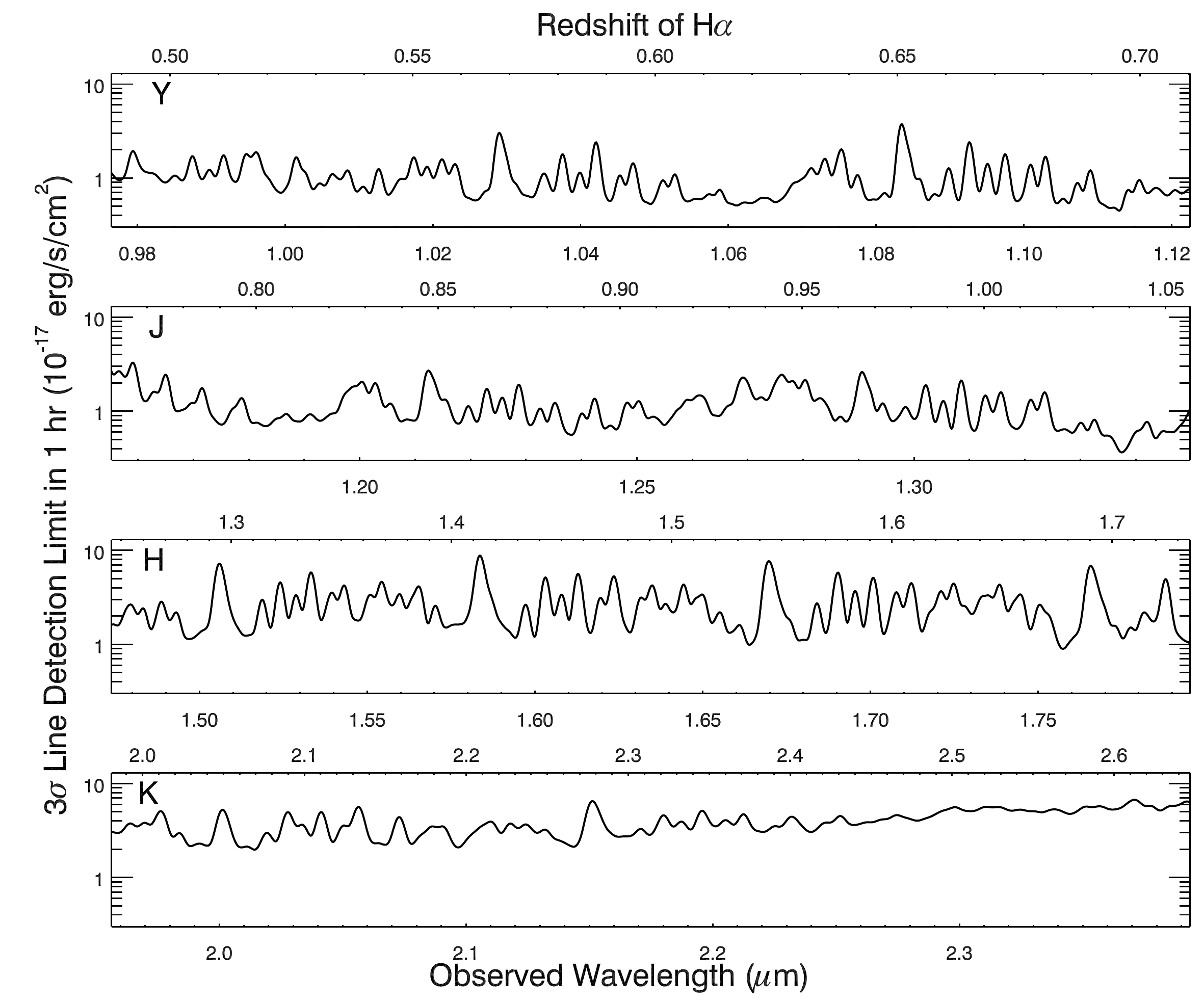}
  \caption{
    \label{fig:linedetect}
    The sensitivity of MOSFIRE for line detection in 1~h in the
    four $YJHK$ filters.  The flux limit is characterized for a
    fiducial \Halpha\ emission line of rest-frame width
    $\sigma_0=85~\kms$ at the $3\sigma$ level, as a function of
    wavelength of the line peak.  We flux calibrated spectra using
    \HST/WFC3 slitless grism observations of a subset of $z\approx1.5$
    galaxies, and so implicitly included a correction for the average
    slit losses.  The 1~h $3\sigma$ flux limit of MOSFIRE is $\sim1
    \times 10^{-17}\ergsex$ in regions with no sky lines in the $YJH$
    filters, and $\sim$4--$6 \times 10^{-17}\ergsex$ in the $K$
    filter.}
\end{figure}

Figure~\ref{fig:linedetect} presents the $3\sigma$ emission line flux
limit in a 1~h exposure, as a function of line-center wavelength
across the $YJHK$ filters.  The 1~h $3\sigma$ flux limit of MOSFIRE is
$\sim1 \times 10^{-17}\ergsex$ in regions with no sky lines in the
$YJH$ filters, and up to ten times shallower in regions with strong
telluric features.  In the $K$ filter, the sensitivity decreases from
$\sim4 \times 10^{-17}\ergsex$ in the blue end to $\sim$$6 \times
10^{-17}\ergsex$ in the red end of the filter.  These Keck/MOSFIRE
emission-line sensitivities are very similar to those reported by the
MOSDEF survey \citep{mosdef}.

\subsection{Comparison with Previous Redshifts}
\label{sxn:compare}

\begin{figure}   
 \epsscale{1.0}
 \plotone{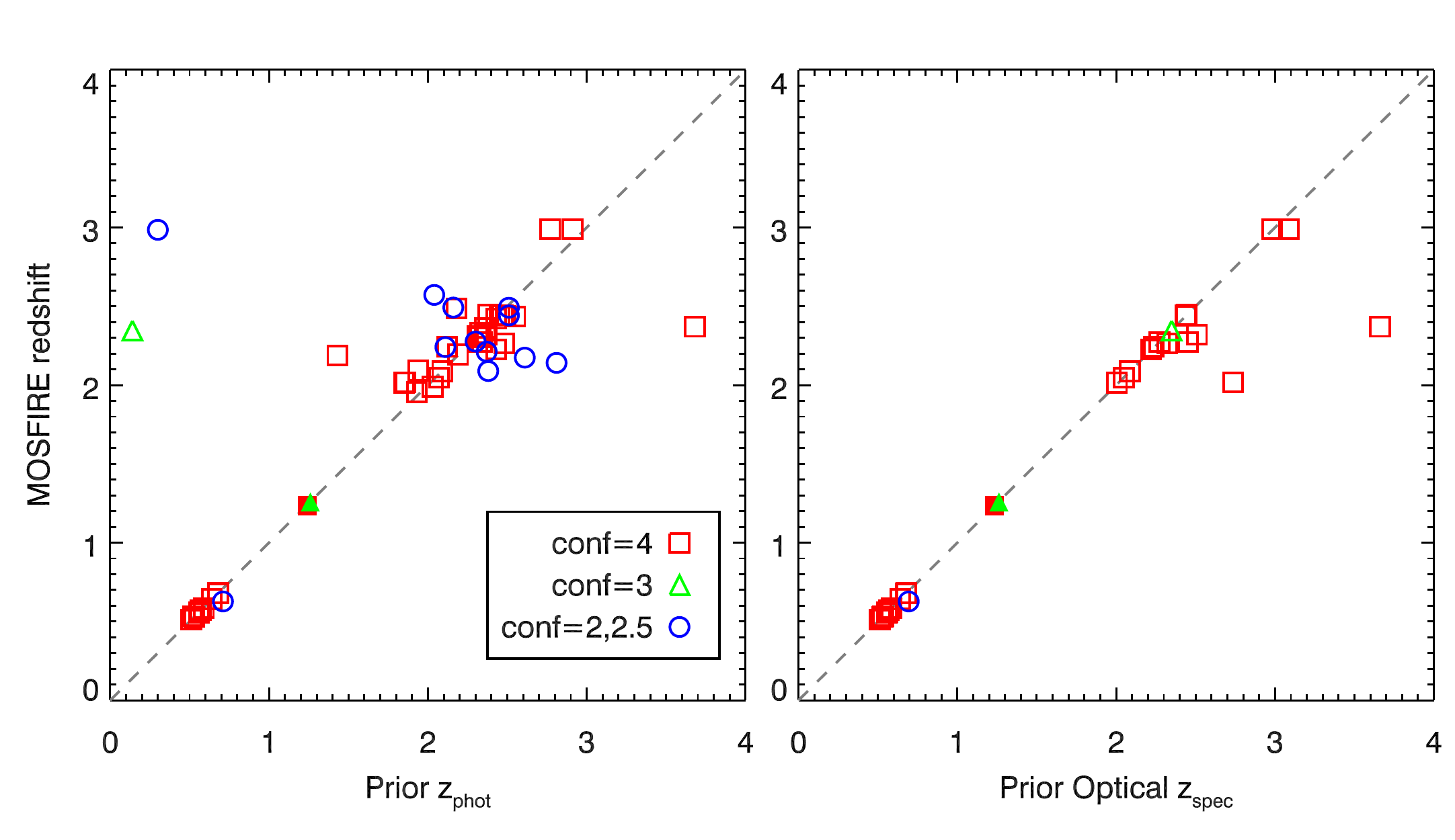}
  \caption{
    \label{fig:comparez}
    A comparison of our MOSFIRE redshifts with prior photometric
    redshifts (left panel) and spectroscopic redshifts from prior
    optical surveys (right panel).  Most galaxies have excellent
    agreement between the MOSFIRE and prior redshifts.  At $2<z<3$,
    prior photometric redshifts and spectroscopic redshifts (based on
    rest-frame UV spectra) are somewhat more uncertain, and the
    rest-frame optical emission lines identified by MOSFIRE result in
    significantly improved redshifts.}
\end{figure}

We compare the MOSFIRE redshifts with the prior photometric redshifts
of our objects in the left panel of Fig.~\ref{fig:comparez}.
Photometric redshifts come from the CANDELS multiwavelength catalog in
GOODS-N (Barro \etal, in preparation), using UV-to-NIR spectral energy
distributions (SEDs) which include 9 \HST\ bands from CANDELS
\citep{candels,candels2} and 24 medium bands from the SHARDS survey
\citep{shards}.  We generated the merged multi-wavelength photometry
following the methods described in \citet{guo13} and \citet{gal13},
accounting for the wavelength-dependent spatial-resolution of the
different imaging.  We used photometric redshifts derived via SED
fitting, selecting the median value from at least 5 different photometric
redshift estimates computed with from a variety of codes (see
\citealt{dah13} for more details).

Most of our photometric redshifts agree well with the spectroscopic
redshifts from MOSFIRE, with the characteristic discrepancy being
$\sigma(\Delta z/z)_\mathrm{NMAD}=0.015$.  The photometric redshift
differs significantly ($\Delta z/z>0.15$) for only 7\% (4/58) of the
$Q_M \ge 3$ sources.

The right panel of Fig.~\ref{fig:comparez} compares the MOSFIRE
redshifts with prior redshifts from optical spectroscopy.  The set of
prior redshifts come from surveys with Keck+DEIMOS \citep{tkrs, bar08,
  coo11}, Keck+LRIS \citep{red06}, and the low-resolution \HST/ACS
G800L grism \citep{fer09}.  In total, 37 MOSFIRE sources with $Q_M \ge
3$ have prior spectroscopic redshifts, and $\sim$90\% (33 of 37) of
these galaxies have nearly identical redshifts (with $|\Delta
z/z|<0.05$).  Only 2 galaxies have spectroscopic redshifts differing
by more than $|\Delta z/z|>0.1$: in both of these cases, the prior
redshifts were of moderate confidence (converted to our scale, $Q=3$),
while the MOSFIRE redshift was based on multiple spectral lines.
Therefore, we conclude that the prior redshifts were incorrect for
these galaxies, with correct redshifts provided by our MOSFIRE
observations.

Seven of our galaxies also have previous near-IR spectroscopy from
Subaru+MOIRCS \citep{yos10}, and we included them in our masks to
measure the relative efficiency of MOSFIRE.
Figure~\ref{fig:moircscompare} compares the signal-to-noise (S/N) of
the \Halpha\ emission line from each instrument, scaled to a 1~h
exposure, with MOIRCS \Halpha\ fluxes and errors derived from Table~3
of \citet{yos10}.  On average, MOSFIRE achieves $\sim$2--3$\times$ higher
emission-line S/N than MOIRCS in the same exposure time, fully
consistent with Keck's 47\% greater collecting area and the
2--5$\times$ throughput advantage of MOSFIRE over MOIRCS.  The scatter
in the S/N comparison is likely due to the differential effects of
telluric features, which tend to have stronger effects on the
lower-resolution MOIRCS observations.

For two objects, the improvement of MOSFIRE is even more dramatic:
these galaxies had emission lines blended with telluric features in
the lower-resolution MOIRCS observations, but our higher-resolution
MOSFIRE spectra resolved the lines.

\subsection{Line Ratios}

Many of our sources have prior redshifts from previous optical
(rest-frame UV) spectroscopy campaigns, but, for most galaxies in the
sample, our MOSFIRE survey provides the first observations of
rest-frame-optical lines.  A number of galaxy properties are encoded
in the strengths and ratios of these optical emission lines.  Here we
investigate galaxy ionization conditions as probed through the ratios
of partially-ionized forbidden lines and Balmer recombination lines in
the ``BPT'' \citep{bpt81} diagram of \Oiii$\lambda$5007/\Hbeta\
vs.\ \Nii$\lambda$6584/\Halpha\ and the ``VO87'' \citep{vo87} diagram
of \Oiii$\lambda$5007/\Hbeta\ vs.\ \Sii$\lambda$(6718+6731)/\Halpha.
Traditionally, the BPT and VO87 diagrams have been used to distinguish
the harder ionizing radiation of active galactic nuclei from typical
star-forming \Hii\ regions \citep[e.g.,][]{kau03, kew06}.  Galaxies at
$z>1$ commonly exhibit high \Oiii/\Hbeta\ ratios compared
to $z \approx 0$ galaxies, causing an offset in the BPT diagram, perhaps
due to higher-ionization \Hii\ regions \citep[e.g.,][]{liu08, bri08,
  kew13, stei14} and/or low-metallicity active galactic nuclei (AGNs)
\citep[e.g.,][]{wri10, tru11, tru13, kew13, jun14}.

\begin{figure}
 \epsscale{1.0}
 \plotone{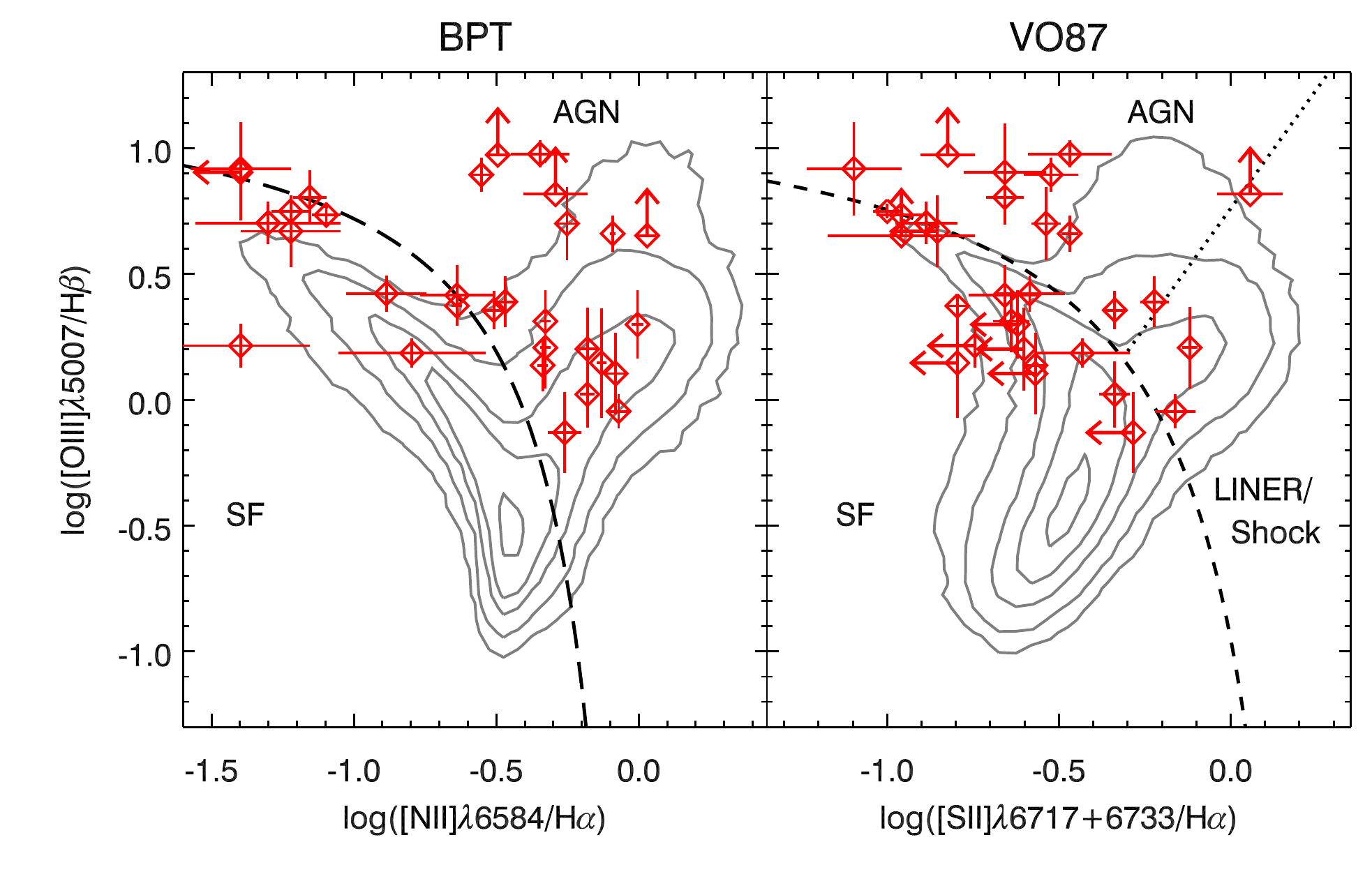}
  \caption{
    \label{fig:bpt}
    The BPT and VO87 line-ratio diagnostic diagrams for the 33 TKRS2
    galaxies with high-confidence ($Q_M \ge 3$) redshifts and
    \Hbeta+\Oiii\ in the $H$ band and \Halpha+\Nii+\Sii\ in the $K$
    band.  For comparison, a sample of $z \approx 0$ SDSS galaxies are
    shown as gray contours \citep[drawn from][]{tru15}.  These TKRS2
    $z \approx 2.3$ galaxies generally have higher \Oiii/\Hbeta\
    ratios than $z \approx 0$ galaxies, as noted in previous work
    \citep[e.g.,][]{tru13, mas14, stei14, coil15}.}
\end{figure}

Figure \ref{fig:bpt} shows our $z \approx 2.3$ TKRS2 galaxies in the
BPT and VO87 diagrams, with a $z<0.1$ sample of SDSS galaxies
\citep[drawn from][]{tru15} shown for comparison.  We measure
rest-frame optical emission lines for our galaxies by subtracting a
local linear continuum and fitting each line with a single Gaussian.
For placement in the BPT or VO87 diagrams, we require at least one
line in each line-ratio pair to be measured at $>1\sigma$: for our
targets, this requires \Oiii\ or \Hbeta\ to be $>1\sigma$-detected in
the $H$-band and \Nii\ or \Halpha\ (for the BPT) or \Sii\ or
\Halpha\ (for the VO87) to be $>1\sigma$-detected in the $K$-band.
Figure~\ref{fig:bpt} shows the 33 galaxies in the $1.99<z<2.49$
redshift interval meeting this criterion for BPT and VO87 line ratios.
Line ratios with only one line measured at $>1\sigma$ are treated as
limits; in our galaxies, this results in \Oiii/\Hbeta\ lower limits
(when \Hbeta\ is not well detected) and \Nii/\Halpha\ or
\Sii/\Halpha\ upper limits (when \Nii\ or \Sii\ is not well detected).

As observed in previous work \citep[e.g.,][]{erb08, tru11, jun14,
  coil15}, we find that our $z \approx 2.3$ targets tend to have
higher \Oiii/\Hbeta\ ratios than $z \approx 0$ galaxies at a fixed
\Nii/\Halpha\ ratio.  In particular, the $z \approx 2.3$ galaxies tend
to lie between the low-metallicity star-forming galaxy locus (upper
left of Fig.~\ref{fig:bpt}) and the AGN locus (upper right).  The $z
\approx 2$ galaxies similarly have high \Oiii/\Hbeta\ ratios in the
VO87 diagram, but have a lower fraction of \Sii/\Halpha\ ratios
elevated above the $z \approx 0$ star-forming galaxy population.
\citet{mas14} similarly noted that $z \approx 2$ emission-line
galaxies are more unusual in the BPT diagram than in the VO87 diagram,
arguing that this results from higher N/O abundance at high redshift
\citep[see also][]{stei14}.  One possible mechanism for producing
higher N/O abundance is Wolf-Rayet stars, which (if present in
sufficient quantities) might also contribute to the higher
\Neiii/\Oiii\ fluxes observed in $z \approx 2$ galaxies \citep{zei14}.
It is not, however, immediately clear why high-redshift \Hii\ regions
would produce a much larger fraction of Wolf-Rayet stars compared to
today.  An alternative explanation for the different BPT and VO87
diagram positions of $z \approx 2$ galaxies is that these galaxies
host a substantial fraction of weak AGNs with low-metallicity
narrow-line regions (NLR) \citep[e.g.,][]{tru11}, since the
\Nii/\Halpha\ ratio is a much more sensitive metallicity indicator
than \Sii/\Halpha.  The AGN NLR gas is located on $\sim$kpc scales,
and so it is quite plausible that it would have the same low
metallicities as typical galaxies at $z \approx 2$.

\section{Summary}

The present survey of the GOODS-North field complements the original
Team Keck Redshift Survey \cite[TKRS,][]{tkrs}, which obtained optical
spectra of galaxies in the same field and helped to establish the
DEIMOS spectrograph and the Keck~II telescope as the world's leading
combination for visible-wavelength multi-object optical spectroscopy
of distant galaxies.  Similarly, the TKRS2 program demonstrates the
unique capabilities of MOSFIRE and the Keck~I telescope to study
sizeable samples of galaxies in the near-IR.  In the spirit of the
original TKRS project, we offer all data products related to the TKRS2
survey --- including raw images, data reduction scripts, 2-D reduced
spectra, and extracted 1-D spectra --- freely to the community to allow
researchers access to representative MOSFIRE spectra.

\acknowledgments

We thank the anonymous referee for helpful comments, and WMKO
Observing Assistants Joel Aycock, Carolyn Jordan, Jason McIlroy, and
Cynthia Wilburn for skillfully operating the Keck~I telescope during
these observations.  GDW is grateful for the many wonderful members of
the WMKO staff who served as valued colleagues and mentors during his
astronomical career, as well as the talented and inspiring community
of Keck observers he was privileged to support in their observing
runs.

UC astronomers acknowledge support for this project from NSF grant AST
08-08133.  JRT acknowledges support provided by NASA through Hubble
Fellowship grant \#51330 awarded by the Space Telescope Science
Institute, which is operated by the Association of Universities for
Research in Astronomy, Inc., for NASA under contract NAS 5-26555.  The
authors wish to recognize and acknowledge the very significant
cultural role and reverence that the summit of Maunakea has always had
within the indigenous Hawaiian community.  We are grateful to have had
the opportunity to conduct observations from this mountain.

{\it Facilities:} \facility{WMKO (MOSFIRE)}.

\clearpage

\clearpage

\begin{figure}
  \epsscale{1.0}
  \plotone{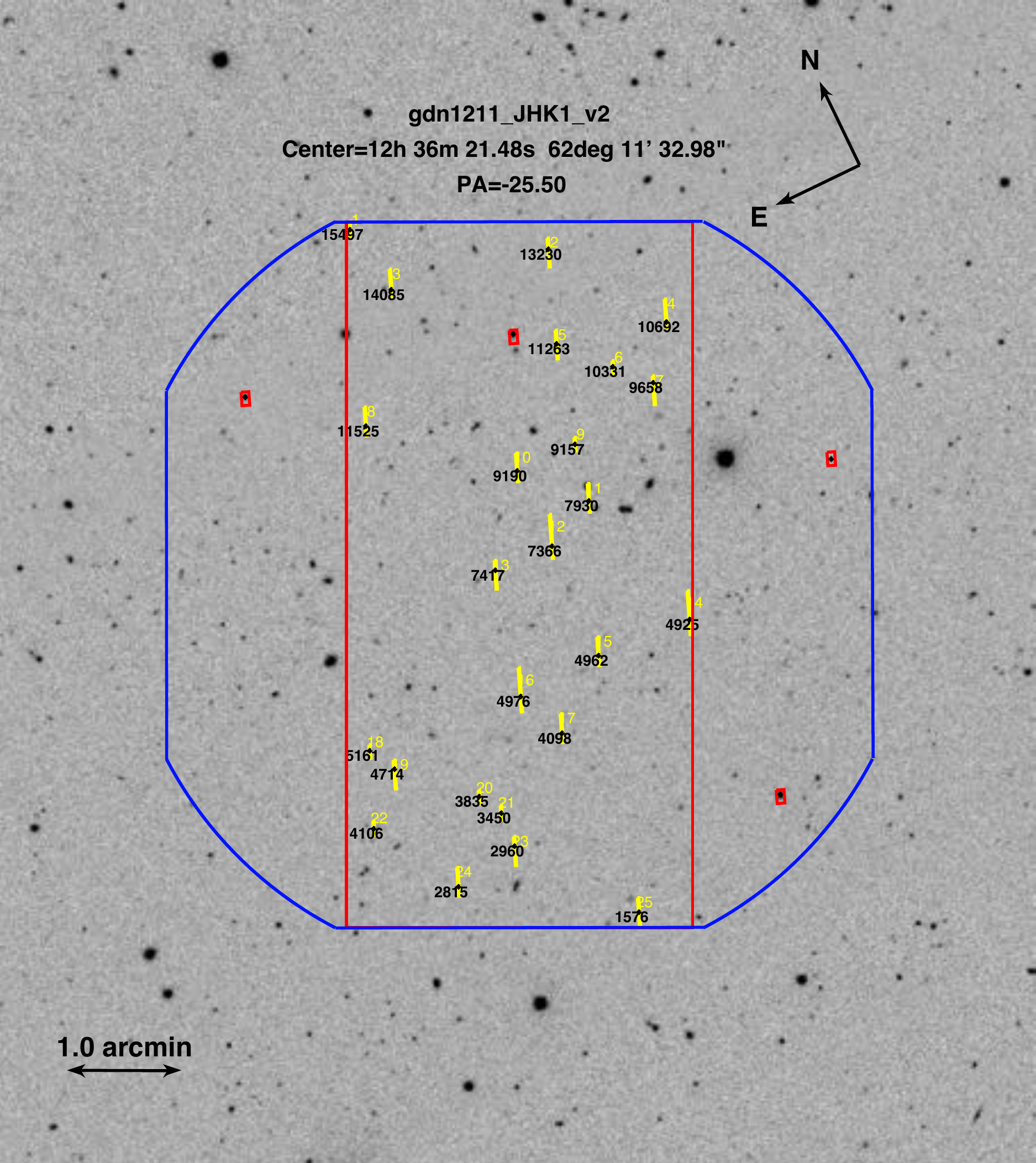}
  \caption{
    \label{fig:gdn1211JHK1}
    Layout for mask \texttt{gdn1211\_JHK1}.  Blue envelope indicates the
    imaging region of MOSFIRE.  Large red rectangle indicates region
    accessible to slits.  Individual slits appear as numbered yellow
    rectangles with corresponding target names as indicated.  Small red
    rectangles indicate mask alignment boxes around stars.  Compass
    rose at top right indicates orientation of the image, and scale
    bar in lower left indicates angular scale.  }
\end{figure}

\begin{figure}
  \plotone{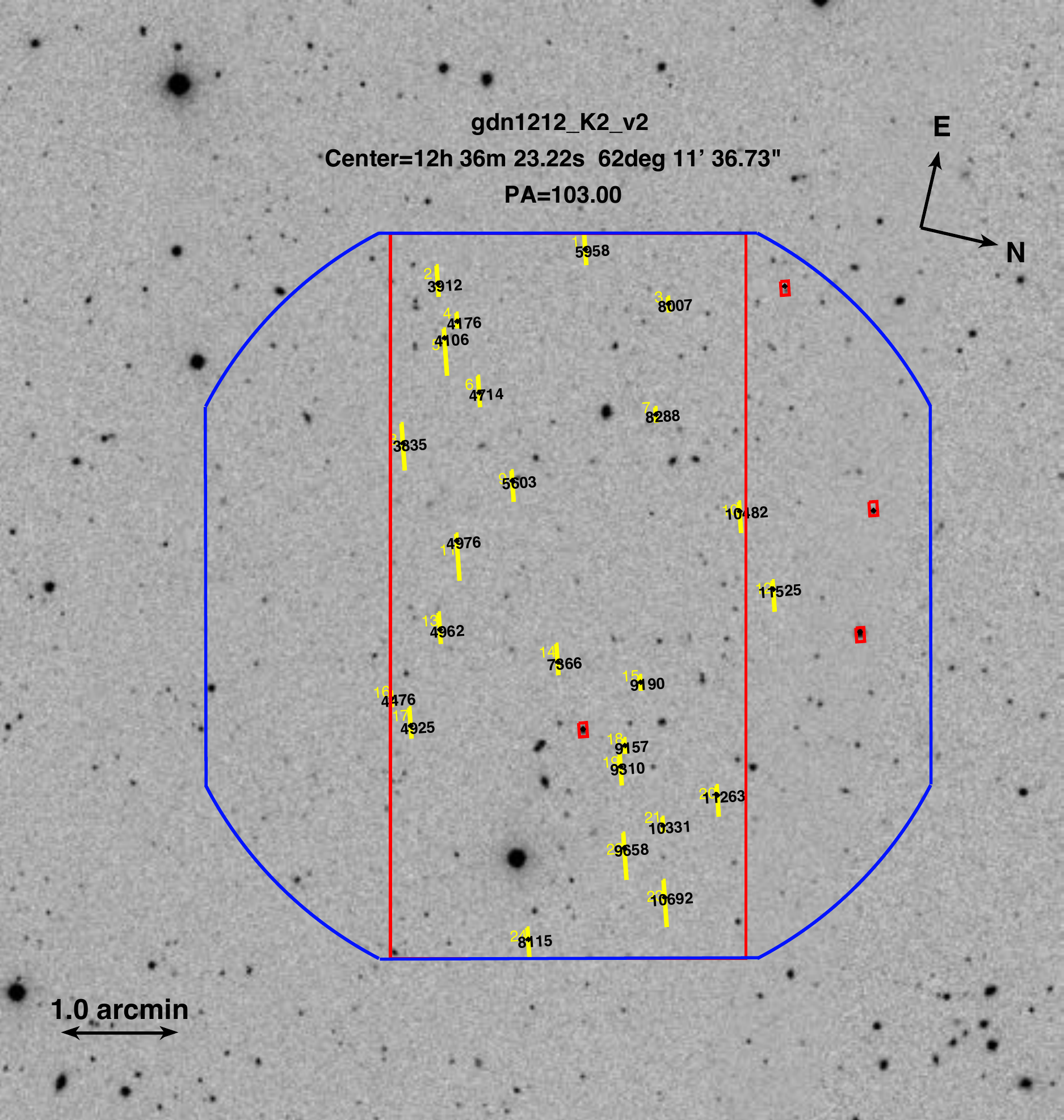}
  \caption{
    \label{fig:gdn1212K2v2}
    Layout for mask \texttt{gdn1212\_K2}; as in Fig.~\ref{fig:gdn1211JHK1}.}
\end{figure}

\begin{figure}
  \plotone{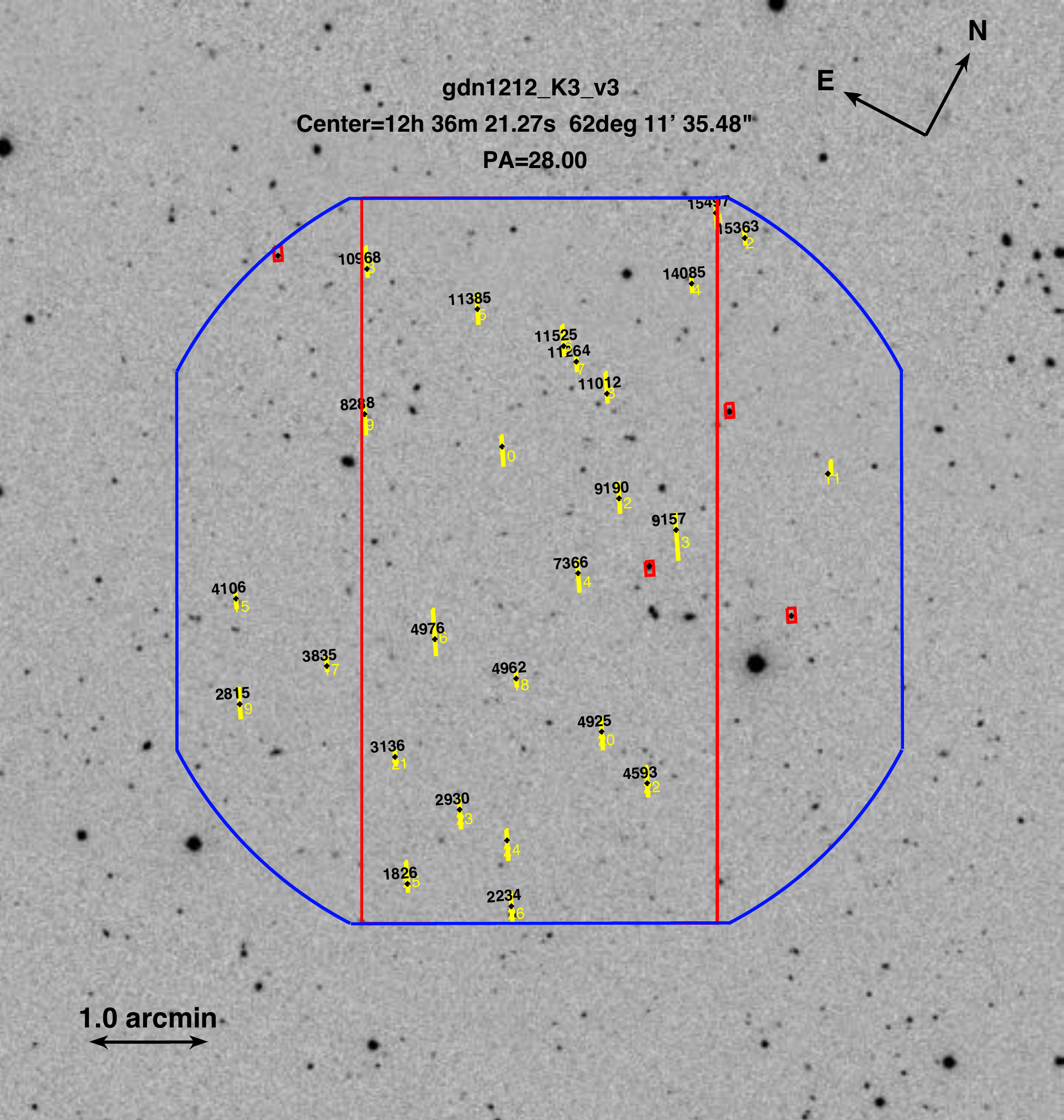}
  \caption{
    \label{fig:gdn1212K3v2}
    Layout for mask \texttt{gdn1212\_K3}; as in Fig.~\ref{fig:gdn1211JHK1}.}
\end{figure}

\begin{figure}
  \plotone{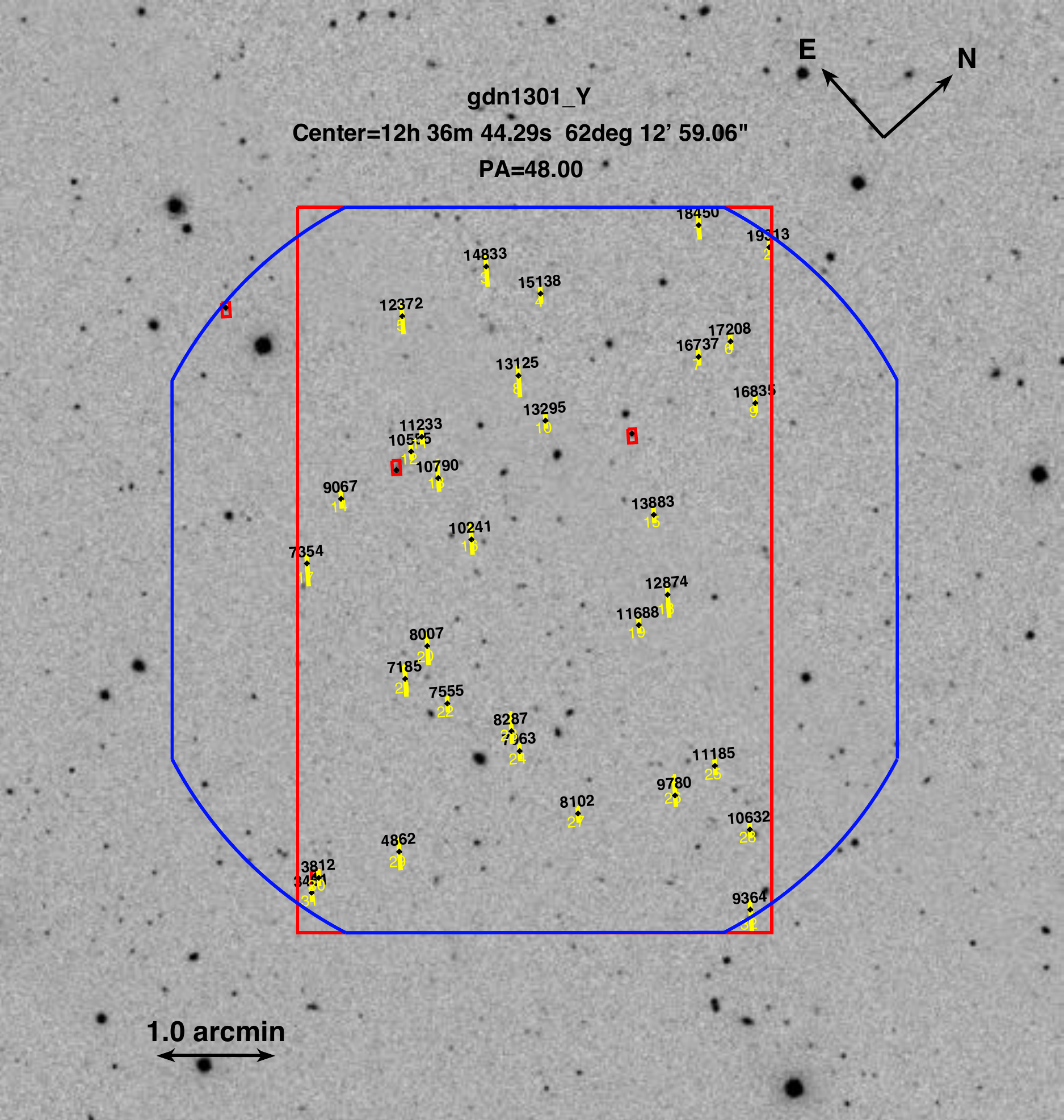}
  \caption{
    \label{fig:gdn1301Y}
    Layout for mask \texttt{gdn1301\_Y}; as in Fig.~\ref{fig:gdn1211JHK1}.}
\end{figure}

\begin{figure}
  \plotone{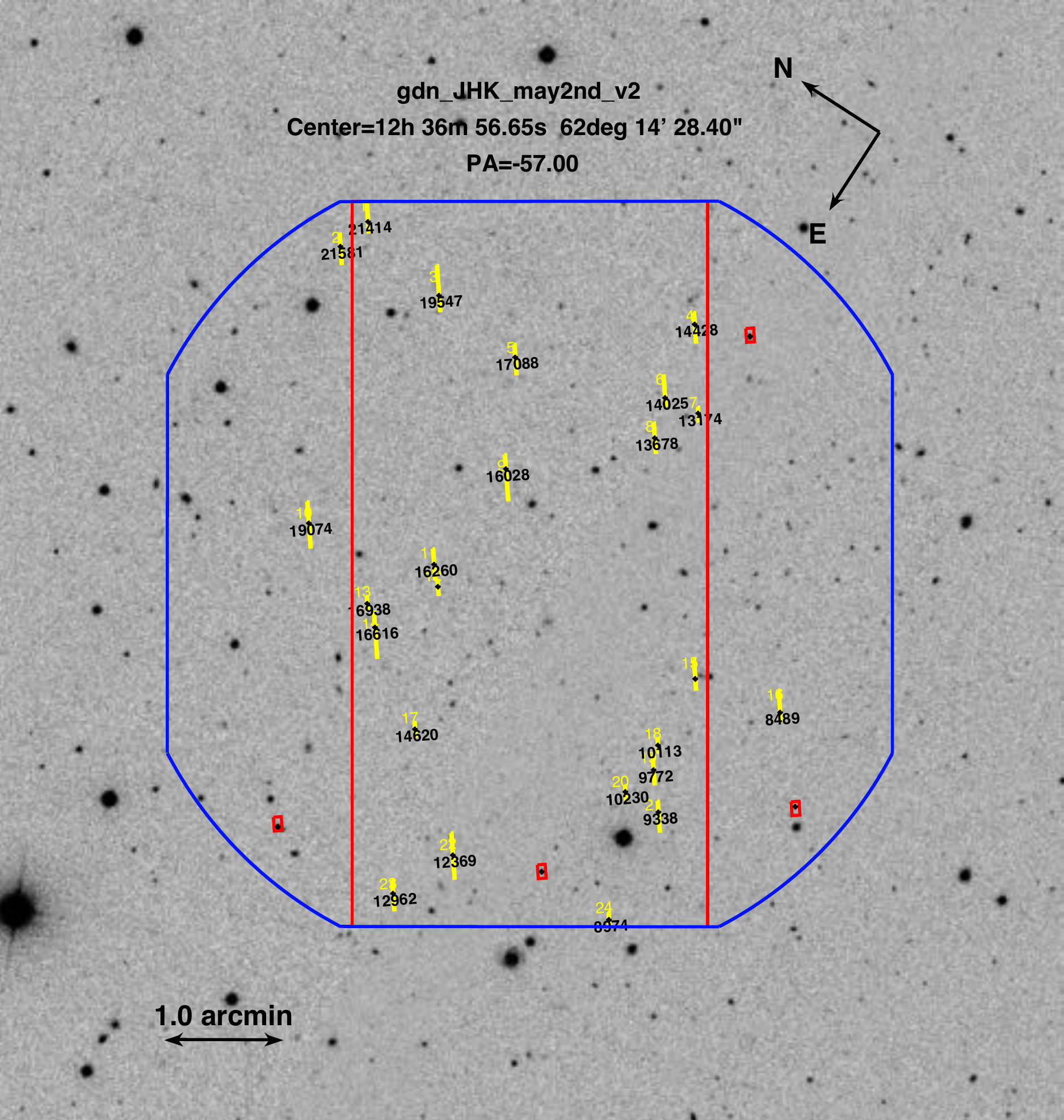}
  \caption{
    \label{fig:gdnJHKmay2nd}
    Layout for mask \texttt{gdn\_JHK\_may2nd}; as in Fig.~\ref{fig:gdn1211JHK1}.}
\end{figure}

\clearpage

\clearpage

\begin{deluxetable}{llrrrrcccc}
  \rotate
  \tablecolumns{10}
  \tablewidth{0pt}
  \tablecaption{
    \label{tab:obs}
    Observation parameters
  }
  \tablehead{
    \colhead{} 
    & \colhead{} 
    & \colhead{} 
    & \colhead{} 
    & \colhead{} 
    & \colhead{} 
    & \multicolumn{4}{c}{Exposure time}
    \\
    \cline{7-10} 
    \colhead{} 
    & \colhead{} 
    & \colhead{$\alpha$\tablenotemark{a}} 
    & \colhead{$\delta$\tablenotemark{a}} 
    & \colhead{P.A.\tablenotemark{b}} 
    & \colhead{} 
    & \colhead{$Y$} 
    & \colhead{$J$} 
    & \colhead{$H$} 
    & \colhead{$K$}  \\
    \colhead{Mask Number} 
    & \colhead{Mask Name} 
    & \colhead{(J2000)} 
    & \colhead{(J2000)} 
    & \colhead{($\degr$)} 
    & \colhead{$N_t$\tablenotemark{c}} 
    & \colhead{(h)} 
    & \colhead{(h)} 
    & \colhead{(h)} 
    & \colhead{(h)}
  } 
  \startdata
  1\dotfill
  & gdn1211\_JHK1 
  & 12:36:21.58
  & +62:11:33.23
  & $-25.5$
  & 25
  & \nodata
  & 1.0
  & 1.0
  & 1.0 \\
  2\dotfill
  & gdn1212\_K2
  & 12:36:23.18
  & +62:11:36.73
  & 103.0
  & 24
  & \nodata
  & \nodata
  & \nodata
  & 2.0 \\
  3\dotfill
  & gdn1212\_K3
  & 12:36:21.27
  & +62:11:35.48
  & 28.0
  & 26
  & \nodata
  & \nodata
  & \nodata
  & 1.5 \\
  4\dotfill
  & gdn1301\_Y
  & 12:36:49.93
  & +62:13:58.06
  & 43.0
  & 32
  & 1.0
  & \nodata
  & \nodata
  & \nodata \\
  5\dotfill
  & gdn\_JHK\_may2nd
  & 12:36:56.73
  & +62:14:27.40
  & $-55.0$
  & 24
  & \nodata
  & 1.0
  & 2.0
  & 3.0 \\
  \enddata
  \tablenotetext{a}{Celestial coordinates of the mask center.}
  \tablenotetext{b}{Projected celestial position angle of the mask; note that the individual MOSFIRE slits are rotated by $+4\degr$ relative to the mask.}
  \tablenotetext{c}{Total number of slits on mask (MOSFIRE allows individual slits to be combined).}
\end{deluxetable}

\clearpage
\begin{deluxetable}{llrl}
  \tablecaption{
    \label{tab:zdist}
    Redshift quality distribution
  }
  \tablehead{
    \colhead{Quality Code} 
    & \colhead{Definition} 
    & \colhead{Number}
    & \colhead{Fraction} 
  }
  \startdata
  4\dotfill & Very secure redshift ($P>99\%$) & 56 & 0.58\\
  3\dotfill & Secure redshift ($P>95\%$) & 2 & 0.02 \\
  2.5\dotfill & Degenerate redshift, but matches $z_\mathrm{phot}$ & 8 & 0.08 \\
  2\dotfill & Degenerate redshift, does not match $z_\mathrm{phot}$ & 4 & 0.04 \\
  1\dotfill & Highly uncertain redshift & 7 & 0.07 \\
  0\dotfill & No redshift & 20 & 0.21 \\
  \enddata
\end{deluxetable}

\clearpage
\begin{deluxetable}{rrrllllllllrrl}
  \tablewidth{636pt}
  \tabletypesize{\scriptsize}
  \rotate
  \tablecaption{
    \label{tab:zcat}
    Redshift catalog
  }
  \tablehead{
    \colhead{} 
    & \colhead{$\alpha$} 
    & \colhead{$\delta$}
    & \colhead{} 
    & \colhead{} 
    & \colhead{} 
    & \colhead{} 
    & \colhead{} 
    & \colhead{} 
    & \colhead{$t_\mathrm{exp}$}
    & \colhead{} 
    & \colhead{} 
    & \colhead{} \\
    \colhead{ID} 
    & \colhead{(J2000)} 
    & \colhead{(J2000)}
    & \colhead{Class}
    & \colhead{$m_\mathrm{F160W}$}
    & \colhead{prior $z_\mathrm{spec}$}
    & \colhead{$z_\mathrm{spec}$ source}
    & \colhead{$z_\mathrm{phot}$}
    & \colhead{Filters}
    & \colhead{(h)} 
    & \colhead{$N_\mathrm{PA}$} 
    & \colhead{$z_M$} 
    & \colhead{$Q_M$}
    & \colhead{Spectral Features} \\
    \colhead{(1)} 
    & \colhead{(2)} 
    & \colhead{(3)}
    & \colhead{(4)} 
    & \colhead{(5)} 
    & \colhead{(6)} 
    & \colhead{(7)} 
    & \colhead{(8)} 
    & \colhead{(9)} 
    & \colhead{(10)} 
    & \colhead{(11)} 
    & \colhead{(12)} 
    & \colhead{(13)} 
    & \colhead{(14)} 
  }
  \startdata
 1576 & 189.10121960 &  62.14110330 &    moircs & 22.45 & \nodata & \nodata & 2.00 & JHK & 3.0 & 1 &  1.998 &  4.0 & hb,o3 \\
 1826 & 189.07826120 &  62.14436230 &    emline & 23.21 & \nodata & \nodata & 2.36 &   K & 1.5 & 1 &  2.302 &  4.0 & ha,n2,s2 \\
 2234 & 189.04746530 &  62.14840220 &    emline & 22.32 & \nodata & \nodata & 2.30 &   K & 1.5 & 1 & \nodata &  0.0 & \nodata \\
 2815 & 189.14816630 &  62.15564680 &    emline & 23.47 & \nodata & \nodata & 2.36 & JHK & 4.5 & 2 &  2.362 &  4.0 & o2,ha,n2 \\
 2930 & 189.07478850 &  62.15701200 &    emline & 23.69 & \nodata & \nodata & 2.11 &   K & 1.5 & 1 & \nodata &  0.0 & \nodata \\
 2960 & 189.12710660 &  62.15744680 &    emline & 23.67 & \nodata & \nodata & 2.31 & JHK & 3.0 & 1 &  2.311 &  4.0 & o3 \\
 3136 & 189.09941700 &  62.15932020 &    emline & 24.32 & \nodata & \nodata & 1.97 &   K & 1.5 & 1 & \nodata &  0.0 & \nodata \\
 3450 & 189.12699890 &  62.16260147 &    emline & 23.45 & 2.086 & Reddy06   & 2.09 & JHK & 3.0 & 1 &  2.088 &  4.0 & hb,o3 \\
 3461 & 189.15730650 &  62.16280760 &    emline & 23.29 & 0.512 & Barger08  & 0.51 &   Y & 1.0 & 1 &  0.513 &  4.0 & ha,s2 \\
 3812 & 189.15918530 &  62.16485290 &    emline & 20.80 & 0.512 & Wirth04   & 0.51 &   Y & 1.0 & 1 &  0.513 &  4.0 & ha,n2,s2 \\
 3835 & 189.13033780 &  62.16611130 &    emline & 23.37 & \nodata & \nodata & 2.32 & JHK & 6.5 & 3 &  2.302 &  4.0 & o2,ha,n2,s2 \\
 3912 & 189.17999268 &  62.16590118 &    emline & 21.82 & 3.661 & Reddy06   & 3.68 &   K & 2.0 & 1 &  2.370 &  4.0 & o3 \\
 4098 & 189.09874990 &  62.16919680 &    emline & 23.32 & \nodata & \nodata & 2.61 & JHK & 3.0 & 1 &  2.174 &  2.0 & ha \\
 4106 & 189.16402480 &  62.16850060 &    moircs & 22.25 & \nodata & \nodata & 2.34 & JHK & 6.5 & 3 &  2.352 &  4.0 & o2,ha,n2,s2 \\
 4176 & 189.16975470 &  62.16971020 &    emline & 22.87 & \nodata & \nodata & 2.30 &   K & 2.0 & 1 &  2.276 &  2.5 & ha \\
 4476 & 189.05470460 &  62.17253080 &    emline & 23.88 & 2.236 & Reddy06   & 2.12 &   K & 2.0 & 1 &  2.243 &  4.0 & ha,n2 \\
 4593 & 189.02857620 &  62.17261300 &    emline & 21.62 & 2.509 & Reddy06   & 2.37 &   K & 1.5 & 1 &  2.321 &  4.0 & ha,n2 \\
 4714 & 189.15032350 &  62.17497220 &    emline & 23.17 & \nodata & \nodata & 2.31 & JHK & 5.0 & 2 &  2.307 &  4.0 & ha,o2 \\
 4862 & 189.14878440 &  62.17573170 & quiescent & 21.65 & \nodata & \nodata & 1.26 &   Y & 1.0 & 1 & \nodata &  0.0 & \nodata \\
 4925 & 189.04794490 &  62.17603260 &    moircs & 23.94 & \nodata & \nodata & 2.09 & JHK & 6.5 & 3 &  2.245 &  4.0 & ha,n2,s2 \\
 4962 & 189.07818880 &  62.17702260 &    emline & 23.79 & 2.322 & Reddy06   & 2.48 & JHK & 6.5 & 3 &  2.265 &  4.0 & o2,ha,n2,s2 \\
 4976 & 189.10537690 &  62.17655430 &    moircs & 22.22 & \nodata & \nodata & 2.09 & JHK & 6.5 & 3 &  2.084 &  4.0 & ha,n2,s2 \\
 5161 & 189.15475520 &  62.17891140 &    emline & 25.43 & \nodata & \nodata & 2.64 & JHK & 3.0 & 1 & \nodata &  0.0 & \nodata \\
 5603 & 189.12673430 &  62.18227440 &    emline & 24.00 & \nodata & \nodata & 2.38 &   K & 2.0 & 1 &  2.089 &  2.5 & ha \\
 5958 & 189.19959020 &  62.18498830 &    emline & 24.75 & 2.349 & Reddy06   & 0.14 &   K & 2.0 & 1 &  2.346 &  3.0 & ha,n2 \\
 7185 & 189.18647840 &  62.19263560 & quiescent & 24.60 & \nodata & \nodata & 0.65 &   Y & 1.0 & 1 & \nodata &  0.0 & \nodata \\
 7354 & 189.23179980 &  62.19318350 &    emline & 22.38 & 0.558 & Wirth04   & 0.56 &   Y & 1.0 & 1 &  0.559 &  4.0 & ha \\
 7366 & 189.07653870 &  62.19417900 &    moircs & 23.17 & 2.390 & Reddy06   & 2.62 & JHK & 6.5 & 3 &  2.398 &  4.0 & o2,ha,n2 \\
 7417 & 189.09559120 &  62.19458080 &    emline & 23.91 & \nodata & \nodata & 2.40 & JHK & 3.0 & 1 & \nodata &  0.0 & \nodata \\
 7555 & 189.17220320 &  62.19468670 &    emline & 21.43 & 0.585 & Barger08  & 0.59 &   Y & 1.0 & 1 &  0.585 &  4.0 & ha,n2,s2 \\
 7930 & 189.06017860 &  62.19777610 &    emline & 23.16 & 2.221 & Reddy06   & 2.43 & JHK & 3.0 & 1 &  2.227 &  4.0 & ha \\
 7963 & 189.14699090 &  62.19774920 & quiescent & 22.57 & 1.223 & Barger08  & 1.21 &   Y & 1.0 & 1 & \nodata &  0.0 & \nodata \\
 8007 & 189.18925060 &  62.19811130 &    emline & 23.00 & \nodata & \nodata & 2.81 &   K & 2.0 & 1 &  2.141 &  2.0 & ha? \\
 8102 & 189.12134530 &  62.19809840 &    emline & 22.60 & 0.529 & Wirth04   & 0.53 &   Y & 1.0 & 1 &  0.530 &  4.0 & ha,n2 \\
 8115 & 188.99305430 &  62.19879980 &    emline & 23.88 & \nodata & \nodata & 2.28 &   K & 2.0 & 1 & \nodata &  0.0 & \nodata \\
 8287 & 189.15315580 &  62.19891510 &    emline & 20.73 & 0.557 & Ferreras09 & 0.56 &   Y & 1.0 & 1 &  0.556 &  4.0 & ha,n2 \\
 8288 & 189.15584380 &  62.19987360 &    emline & 23.62 & \nodata & \nodata & 2.31 &   K & 3.5 & 2 &  2.103 &  1.0 & ha? \\
 8489 & 189.23250720 &  62.20029800 &    emline & 22.39 & 2.737 & Barger08  & 1.86 & JHK & 6.0 & 1 &  2.017 &  4.0 & ha,n2,s2 \\
 8974 & 189.31296290 &  62.20461920 &    emline & 23.34 & \nodata & \nodata & 2.49 & JHK & 6.0 & 1 & \nodata &  0.0 & \nodata \\
 9067 & 189.23950180 &  62.20293930 &    emline & 19.80 & 0.664 & Ferreras09 & 0.66 &   Y & 1.0 & 1 &  0.670 &  1.0 & ha? \\
 9157 & 189.05649000 &  62.20599320 &    emline & 23.20 & 2.437 & Reddy06   & 2.45 & JHK & 6.5 & 3 &  2.441 &  4.0 & o2,ne3,ha,n2,s2 \\
 9190 & 189.07614090 &  62.20616030 &    emline & 23.61 & 2.441 & Reddy06   & 2.49 & JHK & 6.5 & 3 &  2.439 &  4.0 & ha,n2 \\
 9310 & 189.05000530 &  62.20607770 &    emline & 23.38 & \nodata & \nodata & 2.55 &   K & 2.0 & 1 &  2.434 &  4.0 & ha,n2 \\
 9338 & 189.27751870 &  62.20708550 &    emline & 23.74 & 2.443 & Reddy06   & 2.38 & JHK & 6.0 & 1 &  2.445 &  4.0 & ha \\
 9364 & 189.06492560 &  62.20691180 & quiescent & 21.82 & \nodata & \nodata & 1.00 &   Y & 1.0 & 1 & \nodata &  0.0 & \nodata \\
 9658 & 189.02636730 &  62.20912420 &    emline & 23.01 & \nodata & \nodata & 2.18 & JHK & 5.0 & 2 &  2.487 &  4.0 & o2,ha,n2 \\
 9772 & 189.26833810 &  62.21060800 &    emline & 26.02 & \nodata & \nodata & 1.93 & JHK & 6.0 & 1 &  1.956 &  4.0 & ha,n2,s2 \\
 9780 & 189.10558960 &  62.20981490 &    emline & 21.60 & 0.533 & Barger08  & 0.50 &   Y & 1.0 & 1 &  0.532 &  1.0 & ha \\
10113 & 189.26084480 &  62.21222410 &    emline & 22.96 & \nodata & \nodata & 2.43 & JHK & 6.0 & 1 &  2.421 &  4.0 & ha,n2 \\
10230 & 189.27795300 &  62.21239100 &    emline & 22.17 & \nodata & \nodata & 2.37 & JHK & 6.0 & 1 &  2.212 &  2.5 & ha \\
10241 & 189.20414020 &  62.21275500 &    emline & 23.46 & 0.512 & Barger08  & 0.51 &   Y & 1.0 & 1 &  0.682 &  1.0 & ha \\
10331 & 189.03548710 &  62.21373890 &    emline & 24.31 & \nodata & \nodata & 2.11 & JHK & 5.0 & 2 &  2.241 &  2.5 & ha \\
10482 & 189.13310980 &  62.21439890 &    emline & 23.07 & 2.443 & Reddy06   & 2.47 &   K & 2.0 & 1 &  2.444 &  4.0 & ha \\
10555 & 189.23622490 &  62.21462030 & quiescent & 21.58 & 1.234 & Wirth04   & 1.24 &   Y & 1.0 & 1 &  1.235 &  4.0 & hb,o3 \\
10632 & 189.08267550 &  62.21434780 &    emline & 20.33 & 0.694 & Wirth04   & 0.71 &   Y & 1.0 & 1 &  0.627 &  2.0 & ha \\
10692 & 189.01454620 &  62.21620650 &    emline & 23.94 & \nodata & \nodata & 2.51 & JHK & 5.0 & 2 &  2.491 &  2.5 & ha,o2 \\
10790 & 189.22455130 &  62.21502430 &    emline & 19.84 & 0.641 & Wirth04   & 0.64 &   Y & 1.0 & 1 &  0.642 &  4.0 & ha,n2 \\
10968 & 189.17571390 &  62.21808980 &    emline & 23.88 & \nodata & \nodata & 2.04 &   K & 1.5 & 1 &  2.572 &  2.0 & ha? \\
11012 & 189.09411750 &  62.21837350 &    emline & 23.55 & 2.981 & Reddy06   & 2.77 &   K & 1.5 & 1 &  2.990 &  4.0 & hb,o3 \\
11185 & 189.10440320 &  62.21683960 &    emline & 19.52 & 0.518 & Wirth04   & 0.52 &   Y & 1.0 & 1 & \nodata &  0.0 & \nodata \\
11233 & 189.23715020 &  62.21713110 & quiescent & 20.99 & 1.010 & Ferreras09 & 1.24 &   Y & 1.0 & 1 &  1.049 &  1.0 & o3? \\
11263 & 189.04815160 &  62.22022640 &    emline & 22.80 & \nodata & \nodata & 2.16 & JHK & 5.0 & 2 &  2.492 &  2.5 & ha \\
11264 & 189.10681730 &  62.22033690 &    emline & 24.40 & \nodata & \nodata & 2.51 &   K & 1.5 & 1 &  2.442 &  2.5 & ha \\
11385 & 189.14063940 &  62.22037180 &    emline & 23.26 & \nodata & \nodata & 2.34 &   K & 1.5 & 1 & \nodata &  0.0 & \nodata \\
11525 & 189.11224190 &  62.22162740 &    moircs & 23.73 & \nodata & \nodata & 0.54 & JHK & 6.5 & 3 &  2.398 &  4.0 & ha,n2,s2 \\
11688 & 189.15119800 &  62.22218150 &    emline & 22.14 & 0.679 & Wirth04   & 0.68 &   Y & 1.0 & 1 &  0.680 &  4.0 & ha \\
12369 & 189.32200623 &  62.22800064 &    emline & 27.55 & \nodata & \nodata & 0.30 & JHK & 6.0 & 1 &  2.985 &  2.0 & o3 \\
12372 & 189.26808890 &  62.22644260 & quiescent & 20.84 & 1.242 & Barger08  & 1.24 &   Y & 1.0 & 1 &  1.238 &  3.0 & hb,o3? \\
12874 & 189.15210960 &  62.22802410 &    emline & 20.12 & 0.556 & Wirth04   & 0.56 &   Y & 1.0 & 1 &  0.557 &  4.0 & ha,n2 \\
12962 & 189.34192260 &  62.23206070 &    emline & 26.98 & \nodata & \nodata & 0.95 & JHK & 6.0 & 1 & \nodata &  0.0 & \nodata \\
13174 & 189.17022100 &  62.23287270 &    emline & 23.71 & 3.087 & Reddy06   & 2.91 & JHK & 6.0 & 1 &  2.990 &  4.0 & hb,o3 \\
13230 & 189.03771860 &  62.23306260 &    emline & 22.96 & 2.048 & Reddy06   & 2.07 & JHK & 3.0 & 1 &  2.048 &  4.0 & o3 \\
13295 & 189.21578180 &  62.23163490 &    emline & 20.01 & 0.556 & Wirth04   & 0.56 &   Y & 1.0 & 1 &  0.557 &  4.0 & ha,n2,s2 \\
13678 & 189.18366540 &  62.23613790 &    emline & 23.92 & 2.273 & Reddy06   & 2.31 & JHK & 6.0 & 1 &  2.273 &  4.0 & ha \\
13883 & 189.17278880 &  62.23411310 &    emline & 19.75 & 0.555 & Ferreras09 & 0.55 &   Y & 1.0 & 1 & \nodata &  0.0 & \nodata \\
14025 & 189.17174480 &  62.23797870 &    emline & 23.08 & \nodata & \nodata & 2.03 & JHK & 6.0 & 1 &  1.989 &  4.0 & ha,n2 \\
14085 & 189.08706410 &  62.23762110 &    emline & 22.22 & \nodata & \nodata & 1.94 & JHK & 4.5 & 2 &  2.092 &  4.0 & o3,ha,n2,s2 \\
14428 & 189.14829020 &  62.24000410 &    emline & 21.25 & 2.005 & Reddy06   & 1.85 & JHK & 6.0 & 1 &  2.015 &  4.0 & ha,n2,s2 \\
14620 & 189.29724600 &  62.24215960 &    emline & 24.29 & \nodata & \nodata & 1.43 & JHK & 6.0 & 1 &  2.188 &  4.0 & ha,n2 \\
14833 & 189.26218500 &  62.23988580 &    emline & 19.39 & 0.511 & Wirth04   & 0.51 &   Y & 1.0 & 1 &  0.512 &  4.0 & ha,n2 \\
15138 & 189.24516100 &  62.24303100 &    emline & 19.58 & 0.676 & Wirth04   & 0.68 &   Y & 1.0 & 1 &  0.678 &  4.0 & ha,n2 \\
15363 & 189.07931690 &  62.24679590 &    emline & 22.78 & \nodata & \nodata & 2.36 &   K & 1.5 & 1 &  2.307 &  4.0 & ha,n2,s2 \\
15497 & 189.09058260 &  62.24801280 &    moircs & 22.97 & 2.204 & Reddy06   & 2.40 & JHK & 4.5 & 2 &  2.207 &  4.0 & o2,ha,n2 \\
16028 & 189.21590160 &  62.25131100 &    emline & 23.23 & \nodata & \nodata & 2.19 & JHK & 6.0 & 1 &  2.194 &  4.0 & ha,n2 \\
16260 & 189.25190810 &  62.25246110 &    emline & 22.58 & \nodata & \nodata & 2.33 & JHK & 6.0 & 1 &  2.330 &  4.0 & ha,n2 \\
16616 & 189.27755100 &  62.25461080 &    emline & 23.35 & \nodata & \nodata & 2.23 & JHK & 6.0 & 1 &  2.188 &  1.0 & ha? \\
16737 & 189.19911340 &  62.25358100 &    emline & 20.95 & 0.533 & Wirth04   & 0.52 &   Y & 1.0 & 1 &  0.534 &  4.0 & ha,n2 \\
16835 & 189.17730510 &  62.25515580 &    emline & 21.47 & 0.533 & Wirth04   & 0.53 &   Y & 1.0 & 1 &  0.533 &  4.0 & ha,n2,s2 \\
16938 & 189.27283970 &  62.25734420 &    emline & 23.46 & \nodata & \nodata & 2.08 & JHK & 6.0 & 1 & \nodata &  0.0 & \nodata \\
17088 & 189.18602430 &  62.25877660 &    emline & 22.50 & 2.453 & Barger08  & 2.34 & JHK & 6.0 & 1 &  2.273 &  4.0 & ha,n2,s2 \\
17208 & 189.19601500 &  62.25835770 &    emline & 22.71 & 0.570 & Wirth04   & 0.57 &   Y & 1.0 & 1 &  0.571 &  4.0 & ha \\
18450 & 189.22856060 &  62.26592220 &    emline & 21.33 & 0.504 & Barger08  & 0.51 &   Y & 1.0 & 1 &  0.606 &  1.0 & ha? \\
19074 & 189.26210950 &  62.27037430 &    emline & 23.11 & \nodata & \nodata & 2.31 & JHK & 6.0 & 1 & \nodata &  0.0 & \nodata \\
19313 & 189.20938530 &  62.27120450 &    emline & 21.59 & 0.502 & Ferreras09 & 0.50 &   Y & 1.0 & 1 & \nodata &  0.0 & \nodata \\
19547 & 189.18296560 &  62.27247100 &    emline & 21.87 & \nodata & \nodata & 2.36 & JHK & 6.0 & 1 &  2.320 &  4.0 & ha,n2 \\
21414 & 189.17590600 &  62.28647920 &    emline & 22.71 & \nodata & \nodata & 2.62 & JHK & 6.0 & 1 & \nodata &  0.0 & \nodata \\
21581 & 189.18680050 &  62.28775230 &    emline & 22.50 & 2.032 & Reddy06   & 2.09 & JHK & 6.0 & 1 & \nodata &  0.0 & \nodata \\
\enddata
\tablecomments{
  (1)~Object identifier in CANDELS catalog.
(2)~Right ascension of the target, in decimal degrees.
(3)~Declination of the target, in decimal degrees.
(4)~The sample class to which the target belongs, as described in
\S\ref{sxn:selection}.
(5)~The apparent AB magnitude of the target in the F160W
passband \citep{candels2} as measured from CANDELS \HST\ WFC3
imagery.
(6)~The presumed redshift of the target from previous spectroscopic
surveys (when available).
(7)~The source of the presumed redshift.
(8)~The estimated redshift of the source derived via multiband
photometry from the CANDELS survey.
(9)~The list of MOSFIRE passbands in which we observed the target.
(10)~The total MOSFIRE exposure time devoted to the target.
(11)~The number of different position angles at which we observed the target.
(12)~The redshift derived from MOSFIRE spectroscopy.
(13)~The redshift quality code, $Q_M$, as described in \S\ref{sxn:determination}.
(14)~The identification of specific spectral features on which we based the redshift measurement, where ha=\Halpha, hb=\Hbeta, n2=\Nii, ne3=\Neiii, o2=\Oii, o3=\Oiii, s2=\Sii, and question marks indicate marginally-detected lines.
}
\end{deluxetable}

\end{document}